\documentclass[lettersize,journal]{IEEEtran}
\usepackage{amsmath,amsfonts}
\usepackage{algorithmic}
\usepackage{algorithm}
\usepackage{array}
\usepackage[caption=false,font=footnotesize,labelfont=rm,textfont=rm]{subfig}
\usepackage{textcomp}
\usepackage{stfloats}
\usepackage{url}
\usepackage{verbatim}
\usepackage{graphicx}
\usepackage{amssymb}
\usepackage{amsthm}
\usepackage{bm}
\usepackage{multirow}
\usepackage{booktabs}
\usepackage{color}
\usepackage{verbatim}

\def\BibTeX{{\rm B\kern-.05em{\sc i\kern-.025em b}\kern-.08em
    T\kern-.1667em\lower.7ex\hbox{E}\kern-.125emX}}
  
\usepackage{balance}
\usepackage{cite}
\allowdisplaybreaks
\newtheorem{lemma}{Lemma}

\DeclareMathOperator*{\argmin}{arg\,min}

\begin{document}

\title{3D Extended Object Tracking by Fusing Roadside Sparse Radar Point Clouds and Pixel Keypoints}

\author{Jiayin Deng, Zhiqun Hu, Yuxuan Xia, Zhaoming Lu, and Xiangming Wen
  \thanks{This work was supported in part by the Key Research and Development Program of China under Grants SQ2022YFB2600041. \textit{(corresponding author: Zhiqun Hu)}}
  \thanks{Jiayin Deng, Zhiqun Hu, Zhaoming Lu, and Xiangming Wen are with the Beijing Laboratory of Advanced Information Networks, Beijing University of Posts and Telecommunications, Beijing 100876, China (e-mail: jiayindeng@bupt.edu.cn; huzhiqun@bupt.edu.cn; lzy0372@bupt.edu.cn; xiangmw@bupt.edu.cn).}
  \thanks{Yuxuan Xia is with the Department of Electrical Engineering, Linköping University, Linköping 58183, Sweden (e-mail: yuxuan.xia@liu.se).}}

\markboth{IEEE Transactions on Signal Processing,~Vol.~xx, No.~xx, ~2023}%
{Shell \MakeLowercase{\textit{et al.}}: A Sample Article Using IEEEtran.cls for IEEE Journals}


\maketitle

\begin{abstract}
Roadside perception is a key component in intelligent transportation systems. In this paper, we present a novel three-dimensional (3D) extended object tracking (EOT) method, which simultaneously estimates the object kinematics and extent state, in roadside perception using both the radar and camera data. Because of the influence of sensor viewing angle and limited angle resolution, radar measurements from objects are sparse and non-uniformly distributed, leading to inaccuracies in object extent and position estimation. To address this problem, we present a novel spherical Gaussian function weighted Gaussian mixture model. This model assumes that radar measurements originate from a series of probabilistic weighted radar reflectors on the vehicle's extent. Additionally, we utilize visual detection of vehicle keypoints to provide additional information on the positions of radar reflectors. Since keypoints may not always correspond to radar reflectors, we propose an elastic skeleton fusion mechanism, which constructs a virtual force to establish the relationship between the radar reflectors on the vehicle and its extent. Furthermore, to better describe the kinematic state of the vehicle and constrain its extent state, we develop a new 3D constant turn rate and velocity motion model, considering the complex 3D motion of the vehicle relative to the roadside sensor. Finally, we apply variational Bayesian approximation to the intractable measurement update step to enable recursive Bayesian estimation of the object's state. Simulation results using the Carla simulator and experimental results on the nuScenes dataset demonstrate the effectiveness and superiority of the proposed method in comparison to several state-of-the-art 3D EOT methods.
\end{abstract}


\section{Introduction}
\IEEEPARstart{W}{ith} the rapid development of connected vehicles and vehicle-to-everything communications technologies, an increasing number of researches have been conducted to improve sensing range and accuracy by leveraging roadside perception \cite{9477198, 10081379}. As a critical component in this area, object tracking improves object positioning accuracy by leveraging multi-frame data \cite{9868055, 10163834}. Moreover, with the enhancement of sensor resolution, modern sensors can capture multiple measurements from a single object, enabling inference of both its extent (\emph{e.g.}, shape and size) and kinematic (\emph{e.g.}, position and velocity) state.

The mainstream 3D object tracking methods are based on deep learning frameworks. These methods typically involve initial detection of objects' 3D sizes and positions, followed by association with existing objects through a trajectory management module \cite{mao20223d}. Leveraging high-resolution LiDAR and camera sensors, recent fusion-based approaches have demonstrated promising performance in vehicle-mounted perception scenarios \cite{wang2023camo, li2023poly, guo20223d}. However, LiDAR and camera sensors have limited robustness and efficacy in long-distance detection. To address these challenges, there has been growing interest in fusing high-resolution 4D radar with cameras \cite{lei2023hvdetfusion, xiong2023lxl, zheng2023rcfusion, kim2023crn}. Nonetheless, deep learning approaches often require extensive and high-quality annotations, resulting in higher labor costs. Moreover, these methods exhibit poor transferability and interpretability, which can potentially lead to security concerns arising from anomalous outputs.


To construct a model with robust interpretability and meet the demands for high-quality perception using modern high-resolution radar sensors, extended object tracking (EOT) is emerging as a promising technology, which constructs probability models between radar measurements and the object's extent, enabling simultaneous estimation of the kinematic and the extent state of the object \cite{granstrom2022tutorial, granstrom2016extended}. However, due to the object's self-occlusion, the measurements could hardly be considered as Gaussian or uniformly distributed, posing a great challenge to EOT \cite{haag2018radar}.




To address this challenge, recent studies manually divide the vehicle shape into multiple regions with various reflection centers and noise variances. Cao \emph{et al.} \cite{cao2021automotive} proposed an EOT method employing a radar measurement model comprising five parts with fixed relative positions to characterize the point cloud distribution surrounding a rectangular-shaped vehicle. Zhang \emph{et al.} \cite{zhang2021tracking} designed a conditional Gaussian mixture radar measurement model consisting of four components with dynamically updating positions and variances based on the point cloud distribution during tracking. Tuncer \emph{et al.} \cite{tuncer2022multi} proposed a multi-ellipsoid-based EOT model capable of dynamically adjusting the number and position of the ellipsoids. The above methods using manually designed measurement models rely on exploiting the object extent information captured in the point clouds, and thus their performance may drop sharply in the case of sparse radar measurements.

To solve the EOT problem with sparse radar measurements, several studies have focused on pre-learning radar measurement models from extensive data to better depict the real-world characteristic of radar point clouds. Scheel \emph{et al.} \cite{scheel2018tracking} presented a multi-vehicle tracking method based on a variational radar model that is learned from actual data using variational Gaussian mixtures. Xia \emph{et al.} \cite{xia2021learning} employed a learned-based hierarchical truncated Gaussian radar measurement model with structural geometry parameters to track vehicles. However, methods based on learned radar models require an abundance of annotated data, which increases labor costs. Furthermore, the trained radar model is limited to the radar used in the training set and may exhibit poor generalization capabilities.

Considering the inherent problems caused by point cloud data, we propose integrating a probability model of visual measurements into the EOT framework. This integration leverages information from both sensors to enhance tracking results for both the extent and kinematic state of the extended object. The contributions of this work are summarized as follows:


\begin{itemize}
  \item We propose a novel radar measurement model called spherical Gaussian function weighted Gaussian mixture model (SGW-GMM). This model effectively captures the non-uniform distribution of radar measurements and allows for real-time parameter updates based on actual radar measurements.
  \item We propose an elastic skeleton (ES) fusion mechanism, which is integrated into the Bayesian filtering framework. This mechanism creates a virtual spring-damping system that facilitates interaction between radar reflectors and the vehicle's extent. The radar reflectors contribute to refining the vehicle's position, thereby enhancing the accuracy of its kinematic state. Additionally, the vehicle's extent provides prior positions of radar reflectors.
  \item A simple yet effective 3D constant turn rate and velocity (CTRV) motion model based on quaternion algebra is proposed to describe the vehicle's maneuvering characteristics. In addition, due to the more realistic continuous motion pattern, the proposed model imposes constraints on the vehicle's extent.
  \item Variational approximation is adopted to enable a tractable measurement update step in recursive Bayesian estimation, allowing for accurate extent and kinematic state tracking of the vehicle. Simulation results using the Carla simulator demonstrate the significant performance improvement of our proposed method compared to existing state-of-the-art 3D EOT algorithms.
\end{itemize}

This paper follows specific notational conventions: scales are represented with lowercase italic letters, \emph{e.g.}, $x$, vectors with bold lowercase letters, \emph{e.g.}, $\mathbf{x}$, and matrices with bold uppercase letters, \emph{e.g.}, $\mathbf{X}$. Additionally, $\mathbf x^\text{T}$ and $\mathbf X^\text{T}$ denote the transpose of vector $\mathbf x$ and matrix $\mathbf X$, respectively; $\propto$ indicates equality up to a normalization factor; $\mathcal N \left( \mathbf x; \bm \mu, \bm \Sigma \right)$ denotes the Gaussian probability density function PDF (of random vector x) with mean $\bm \mu$ and covariance matrix $\bm \Sigma$. The norm of vector $\mathbf v$ is denoted $\| \mathbf v \|$. The tilde $(\tilde{\cdot})$ above a variable indicates that it is a quaternion or the quaternion form of the vector.

The paper is organized as follows. Section \uppercase\expandafter{\romannumeral2} provides preliminary information about quaternions and coordinate systems. In section \uppercase\expandafter{\romannumeral3}, we introduce the system models for our proposed 3D EOT method. Section \uppercase\expandafter{\romannumeral4} presents the implementation of inference using the variational approximation. The simulation and experimental results are outlined in section \uppercase\expandafter{\romannumeral5}. Finally, we conclude and present future works in section \uppercase\expandafter{\romannumeral6}.

\section{Quaternions and Coordinate Systems}

In this section, we first introduce quaternion and its relationships with corresponding rotation vector and matrix. Then, we delve into the coordinate systems employed in this paper and explain the transformations between them.

\subsection{Quaternion}

Quaternion is a widely used mathematical tool for compactly representing non-singular 3D rotation transformation. A quaternion is commonly expressed as a four-dimensional vector \cite[Eq. 7]{sola2017quaternion}:
\begin{equation}
  \tilde{\mathbf q} \triangleq \begin{bmatrix} q_w & \mathbf q_v^{\text{T}} \end{bmatrix}^\text{T} = \begin{bmatrix} q_w & q_x & q_y & q_z \end{bmatrix}^\text{T} \text{ , }
\end{equation}
where $q_w$ refers to the real part and $\mathbf q_v$ is the imaginary part.

Rotation using quaternions is equivalent to multiplying the vector by a rotation matrix. For instance, the rotation from a 3D vector $\mathbf v$ to $\mathbf v'$, can be expressed using either a quaternion or a rotation matrix as:
\begin{equation}
  \label{eqn:rotation using quaternion}
  \tilde{\mathbf v}'=\tilde{\mathbf q} \odot \tilde{ \mathbf v } \odot \tilde{\mathbf q}^* \text{ } \Leftrightarrow \text{ } \mathbf v' = \mathbf R\{ \tilde{\mathbf q} \} \cdot \mathbf v \text{, }
\end{equation}
where $\tilde{\mathbf q}^*= \begin{bmatrix}q_w & -\mathbf q_v^\text{T} \end{bmatrix}^\text{T}$ represents the conjugate of the quaternion. The operators $\odot$ and $\mathbf R\{ \cdot \}$, respectively, denote the product of two quaternions and the $3 \times 3$ rotation matrix form corresponding to the quaternion, defined as:
\begin{align}
  \tilde{\mathbf p} \odot \tilde{\mathbf q} = & \begin{bmatrix} p_wq_w - \mathbf p_v^\text{T} \mathbf q_v \\ p_w \mathbf q_v + q_w \mathbf p_v + \mathbf p_v \times \mathbf q_v \end{bmatrix} \text{, } \\[5pt]
  \mathbf R\{ \tilde{\mathbf q} \} = & \left( q_w^2-\mathbf q_v^\text{T} \mathbf q_v \right) \mathbf I + 2\mathbf q_v \mathbf q_v^\text{T}+2q_w \left[ \mathbf q_v \right]_\times \text{, }
\end{align}
where $\mathbf I$ is a $3 \times 3$ identical matrix, and $[\cdot]_\times$ represents the cross-product matrix of the vector:
\begin{equation}
  \label{eqn: skew matrix}
  \left[ \mathbf q_v \right]_\times = \begin{bmatrix} 0 & -q_z & q_y \\ q_z & 0 & -q_x \\ -q_y & q_x & 0 \end{bmatrix} \text{. }
\end{equation}
Noted that each vector has a corresponding quaternion form, \emph{e.g.} the quaternion form of $\mathbf v$ is $\tilde{\mathbf v} = \begin{bmatrix}0 & \mathbf v^\text{T} \end{bmatrix}^\text{T}$, which is also called pure quaternion.


To simplify the derivation in the following sections, we introduce an alternative representation of rotation, namely, the rotation vector $\bm \theta$ $(\text{where } \| \bm \theta \| \in [0, 2\pi))$. The corresponding rotation matrix can be expressed as follows:
\begin{align}
  \label{eqn:R_theta}
  \mathbf R\{ \bm \theta \} = & \exp \left( \left[ \bm \theta \right]_\times \right) = \mathbf I + \sum_{i=1}^{\infty} \frac{1}{i!} [\bm \theta]^i_\times \notag \\
  = &\text{ } \mathbf I + \sin \| \bm \theta \| \left[\frac{\bm \theta}{\|\bm \theta\|}\right]_\times + \left( 1 - \cos \| \bm \theta \| \right) \left[\frac{\bm \theta}{\|\bm \theta\|}\right]^2_\times \text{. } 
\end{align}
Notably, if the quaternion $\tilde{\mathbf q}$ and the rotation vector $\bm \theta$ represent the same rotation transformation, their corresponding rotation matrices are identical, i.e., $\mathbf R\{\tilde{\mathbf q}\} = \mathbf R\{\bm \theta\}$. In this case, there is a one-to-one correspondence between $\tilde{\mathbf q}$ and $\bm \theta$, and to specifically denote these correspondences, we introduce the following symbols as in \cite[Eq. 101, 103]{sola2017quaternion}
\begin{equation}
  \label{eqn: Exp and Log}
  \tilde{\mathbf q} = \mathrm{Exp}\left( \bm \theta \right) \text{ and }
  \bm \theta = \mathrm{Log}\left( \tilde{\mathbf q} \right) \text{. }
\end{equation}

\subsection{Coordinate Systems}


Our research scenario focuses on the roadside integrated radar and camera sensor, with the camera coordinate system (CCS) axis aligned with the radar coordinate system (RCS) axis, and their origins are almost coincident. Therefore, the CCS and RCS can be considered the same, see also \cite{scholler2019targetless}, and we refer to this unified system as the sensor coordinate system (SCS). As for the vehicle coordinate system (VCS), the x-axis, y-axis, and z-axis point forward, leftward, and upward of the vehicle, respectively. The origin of the VCS is defined as the midpoint at the bottom of the vehicle.

According to the definition of the coordinate systems and quaternion, a 3D point in the VCS can be transformed to the image pixel coordinate system (PCS) through two steps. Specifically, to transform the position of the $t$-th radar reflector $(\mathbf u_t)$ in VCS:

We first transform $\mathbf u_t$ from VCS to SCS through rotation and translation, to obtain the coordinate $\mathbf u_t^\mathcal{S}$:
\begin{equation}
  \label{eqn:rigid transformation}
  \mathbf u_t^\mathcal{S} = \mathbf R\{ \bm \theta \} \cdot \mathbf u_t + \mathbf p \text{, } 
\end{equation}
where $\bm \theta$ and $\mathbf p$ are the vehicle's pose and position, respectively. Then, we project the point from SCS to the PCS, to obtain the pixel coordinates $\mathbf u_t^\mathcal{I}$ using the distortion-free pinhole camera model \cite{zhang2000flexible}:
\begin{subequations}
  \label{project transformation}
  \begin{gather}
    \kappa \cdot \begin{bmatrix} \left( \mathbf u_t^\mathcal{I} \right)^\text{T} & 1 \end{bmatrix}^\text{T} = \mathbf K \cdot \mathbf u_t^\mathcal{S} \text{, } \\
    \mathbf K = \begin{bmatrix} \frac{f}{d_x} & 0 & u_0 \\ 0 & \frac{f}{d_y} & v_0 \\ 0 & 0 & 1 \end{bmatrix} \text{, }
  \end{gather} 
\end{subequations}
where $\kappa$ is a scale factor, and the matrix $\mathbf K$ is the camera's intrinsic parameters matrix, comprising camera focus $(f)$, parameters of the photosensitive chip $(d_x, d_y)$, and the center of the image $(u_0, v_0)$.





\section{System Model}

\begin{figure}[tbp]
  \centering
  \includegraphics[width=0.48\textwidth]{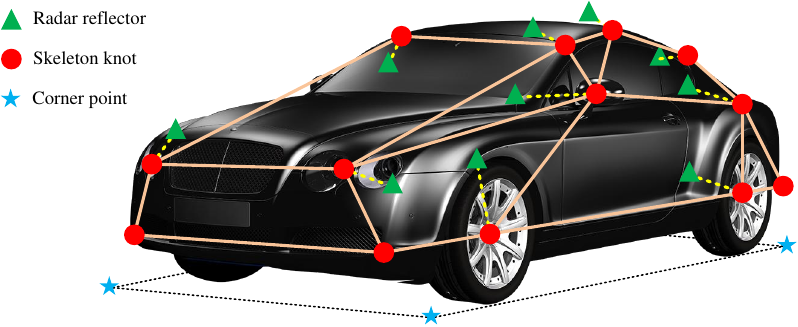}
  \vspace{5pt}
  \caption{Illustration of 13 of the vehicle's skeleton knots and their corresponding radar reflectors, along with 3 of the corner points. The yellow dashed lines represent predefined associations between skeleton knots and reflectors. Visual measurements are named keypoints which consist of pixel coordinates of skeleton knots and corner points, while radar measurements are 3D point clouds generated by radar reflectors based on the probabilities.}
  \label{fig:car skeleton}
\end{figure}



We assume the vehicle moves in a 3D space, and the vehicle's extent is represented by its length, width, and skeleton. As shown in Fig.~\ref{fig:car skeleton}, the skeleton consists of $T$ components, each of which contains a skeleton knot and a radar reflector. The state of the $t$-th component is denoted as $\bm \vartheta_t$ and includes the position of the reflector $(\mathbf u_t)$, the position of the knot $(\bm \varpi_t)$, and the velocity of the knot $(\mathbf v_t)$:
\begin{equation}
  \label{skeleton state}
  \bm \vartheta_t = \begin{bmatrix} \mathbf u_t^\text{T} & \bm \varpi_t^\text{T} & \mathbf v_t^\text{T} \end{bmatrix}^\text{T} \text{. } 
\end{equation}

The vehicle's kinematic state is defined as:
\begin{equation}
  \label{kinematic state}
  \mathbf x = \begin{bmatrix} \mathbf p^\text{T} & v & \bm \theta^\text{T} & \bm \omega^\text{T} & \bm \xi^\text{T} \end{bmatrix}^\text{T} \text{, } 
\end{equation}
where $\mathbf{p}$ represents the 3D position of the vehicle, $v$ is its speed, $\bm{\theta}$ denotes the pose in terms of a rotation vector, $\bm{\omega}$ signifies the angular velocity, $\bm{\xi} = \begin{bmatrix} l & w \end{bmatrix}^\text{T}$ represents the length and width of the vehicle's bottom. Here, the quaternion form of $\bm{\theta}$ is denoted as $\tilde{\mathbf{q}} = \mathrm{Exp}\left( \bm{\theta} \right)$.

\subsection{The 3D CTRV Motion Model} \label{sec:The 3D CTRV Motion Model}

The 3D CTRV motion model in vehicle-mounted perception \cite{guo20223d} assumes that the vehicle primarily changes in azimuth relative to the sensor, while the pitch and roll angles are always 0, whereas our research focuses on roadside perception, where the vehicle's orientation relative to the sensor includes unknown azimuth, pitch, and roll angles. 

In these contexts, we extend the 2D CTRV model \cite{li2003survey} into 3D space by using quaternion-based descriptions of 3D rotation \cite[Eq. 201]{sola2017quaternion} to model all three angles of the vehicle. The time derivative of the 3D CTRV kinematic state is formulated as:
\begin{subequations}
  \label{true kinematic model}
  \begin{gather}
    \dot{\mathbf p} = v\mathbf d \text{, } \label{eqn: dot_p} \\
    \frac{\mathrm{d}\left( v\mathbf d \right)}{\mathrm{d}t}=\dot{v} \mathbf d+v \dot{\mathbf d} \text{, } \\
    \dot{v}=n_v \text{, } \\
    \tilde{\mathbf d}=\tilde{\mathbf q} \odot \tilde{\mathbf u}^d \odot \tilde{\mathbf q}^* \text{, } \\
    \dot{\tilde{\mathbf q}}=\frac{1}{2} \tilde{\bm \omega} \odot \tilde{\mathbf q} \text{, } \label{eqn: dot_q} \\
    \dot{\bm \omega}=\mathbf n_\omega \text{, } \\
    \dot{\bm \xi} = \mathbf n_{\xi} \text{, }
  \end{gather} 
\end{subequations}
where $\mathbf d$ is a unit vector represents the direction of velocity, and the quaternion form of vectors $\mathbf u^d$, $\mathbf d$ and $\bm \omega$ are represented as $\tilde{\mathbf u}^d$, $\tilde{\mathbf d}$ and $\tilde{ \bm \omega }$, respectively. The vector $\mathbf u^d = \begin{bmatrix} 1 & 0 & 0 \end{bmatrix}^\text{T}$ represents the unit direction vector of the forward direction of the vehicle in the VCS. The $n_v$, $\mathbf n_{\xi}$ and $\mathbf n_{\omega}$ are zero-mean Gaussian noises.

To deal with the nonlinearity introduced by quaternion operations, we employ the error state filtering \cite{he2021adaptive}. Specifically, $\mathbf{x} = \mathbf{x}_{ref} \oplus \delta \mathbf{x}$, where $\mathbf{x}_{ref}$ represents a time-varying reference-state without randomness, while $\delta \mathbf{x}$ is a zero-mean random variable, representing the error-state. As a result, the vehicle's motion model described in \eqref{true kinematic model} can be divided into reference-state kinematics and error-state kinematics.

\subsubsection{Reference-state Kinematics}

The motion model of the reference-state can be obtained by replacing the variables in \eqref{true kinematic model} with reference-states:
\begin{subequations}
  \label{reference-state kinematic}
  \begin{align}
    \dot{\mathbf p}_{ref} & = v_{ref} \cdot \mathbf R_{ref} \mathbf u^d \text{, } \label{reference-state p} \\
    \dot{v}_{ref} & = 0 \text{, } \\
    \dot{\tilde{\mathbf q}}_{ref} & = \frac{1}{2} \tilde{\bm \omega}_{ref} \odot \tilde{\mathbf q}_{ref} \text{, } \label{reference-state kin q} \\
    \dot{\bm \omega}_{ref} & = 0 \text{, } \\
    \dot{\bm \xi}_{ref} & = 0 \text{. }
  \end{align} 
\end{subequations}
The time-variant variable $\mathbf R_{ref}$ represents the rotation matrix form of the vehicle's pose, where $\mathbf R_{ref} = \mathbf R\{ \tilde{\mathbf q}_{ref} \}$. Therefore, an equivalent form of \eqref{reference-state kin q} can be rewritten in rotation matrix form, as stated in \cite[Eq. 67]{sola2017quaternion}
\begin{equation}
  \label{eqn: R_ref_derivative}
  \dot{\mathbf R}_{ref} = \left[ \bm \omega_{ref} \right]_\times \mathbf R_{ref} \text{, }
\end{equation}

Let us denote the current time step as $k$ and the time interval between two consecutive steps as $\Delta t$. By integrating \eqref{reference-state kinematic}, we obtain the time update of the reference-state kinematics as follows:
\begin{equation}
  \label{reference-state predict}
  \mathbf x_{ref, k+1} = f \left( \mathbf x_{ref, k} \right) \text{, } 
\end{equation}
with
\begin{subequations}
  \label{eqn:state predict1}
  \begin{align}
    \mathbf p_{ref,k+1} =&\text{ } \mathbf p_{ref,k} + v_{ref,k} \Bigg\{ \mathbf I \Delta t + \frac{1}{\left\| \bm \omega_{ref,k} \right\|} \left[ \frac{\bm \omega_{ref,k}}{\left\| \bm \omega_{ref,k} \right\|} \right]_\times \notag \\
    & \times \left( 1- \cos \left( \left\| \bm \omega_{ref,k} \right\| \Delta t \right) \right) + \left[ \frac{\bm \omega_{ref,k}}{\left\| \bm \omega_{ref,k} \right\|} \right]_\times^2 \notag \\
    & \times \left( \Delta t - \frac{\sin \left( \left\| \bm \omega_{ref,k} \right\| \Delta t \right)}{\left\| \bm \omega_{ref,k} \right\|} \right) \Bigg\} \cdot \mathbf R \{ \tilde{\mathbf q}_{ref,k} \} \mathbf u^d \text{, } \label{eqn:p_ref} \\
    v_{ref,k+1} & = v_{ref,k} \text{, } \label{eqn:v_ref} \\
    \tilde{\mathbf q}_{ref,k+1} & = \mathrm{Exp} \left( \bm \omega_{ref,k} \Delta t \right) \odot \tilde{\mathbf q}_{ref,k} \text{, } \label{eqn:q_ref} \\
    \bm \theta_{ref,k+1} & = \mathrm{Log} \left( \tilde{\mathbf q}_{ref,k+1}\right) \text{, } \label{eqn:theta_ref} \\
    \bm \omega_{ref,k+1} & = \bm \omega_{ref,k} \text{, } \label{eqn:omega_ref} \\
    \bm \xi_{ref,k+1} & = \bm \xi_{ref,k} \text{, } \label{eqn:xi_ref}
  \end{align} 
\end{subequations}
The derivation for obtaining \eqref{eqn:state predict1} is in Appendix A-A.

\subsubsection{Error-state Kinematics}

The error-state motion model is derived by subtracting \eqref{reference-state kinematic} from \eqref{true kinematic model}. To obtain a closed form expression for the predicted object state, we linearize the error-state kinematics using Taylor series and omitting second-order small quantity. This gives:
\begin{equation}
  \label{error-state kinematic}
  \dot{\delta \mathbf x} \approx \mathbf F \delta \mathbf x + \mathbf w \text{, } 
\end{equation}
where
\begin{align}
  \label{eqn: F_M0_M1}
  \delta \mathbf x = \begin{bmatrix} \delta \mathbf p \\ \delta v \\ \delta \bm \theta \\ \delta \bm \omega \\ \delta \bm \xi \end{bmatrix} \text{ , } \mathbf F = \begin{bmatrix} 0 & \mathbf M_0 & \mathbf M_1 & 0 & 0 \\ 0 & 0 & 0 & 0 & 0 \\ 0 & 0 & \mathbf M_2 & \mathbf I & 0 \\ 0 & 0 & 0 & 0 & 0 \\ 0 & 0 & 0 & 0 & 0 \end{bmatrix} \text{ , } \mathbf w = \begin{bmatrix} 0 \\ n_v \\ 0 \\ \mathbf n_\omega \\ \mathbf n_\xi \end{bmatrix} \text{, } \notag \\
  \mathbf M_0 = \mathbf R_{ref} \mathbf u^d \text{ , } \mathbf M_1 = -v_{ref}[\mathbf R_{ref} \mathbf u^d]_\times \text{ , } \mathbf M_2 = [\bm \omega_{ref}]_\times \text{, }
\end{align}
where $\mathbf w \sim \mathcal N \left( 0, \mathbf W \right)$ represents a zero-mean Gaussian noise with covariance $\mathbf W$. The derivation for obtaining \eqref{error-state kinematic} is in Appendix A-B1.

By integrating \eqref{error-state kinematic}, we obtain the time update of the error-state kinematics:
\begin{subequations}
  \label{error-state time update}
  \begin{align}
    \delta \mathbf x_{k+1} & = \Phi_k \delta \mathbf x_k + \mathbf w_{k} \text{, } \label{eqn: delta_x}\\
    \delta \mathbf x_{0}   & \sim \mathcal N \left( 0, \hat{\mathbf P}_{0|0} \right) \text{, }
  \end{align} 
\end{subequations}
where $\mathbf w_{k} \sim \mathcal N \left( 0, \mathbf W \Delta t \right)$, $\hat{\mathbf P}_{0|0}$ is the initialized error-state covariance matrix and $\Phi_k$ is the transition matrix, with the closed form expression:
\begin{equation}
  \label{eqn:error-state transition matrix}
  \Phi = \begin{bmatrix} \mathbf I & \mathbf M_0 \Delta t & \mathbf M_1 \Sigma_1 & \mathbf M_1 \Sigma_2 & 0 \\ 0 & \mathbf I & 0 & 0 & 0 \\ 0 & 0 & \Sigma_0 & \Sigma_1 & 0 \\ 0 & 0 & 0 & \mathbf I & 0 \\ 0 & 0 & 0 & 0 & \mathbf I \end{bmatrix} \text{, } 
\end{equation}
where
\begin{subequations}
  \label{eqn:Sigma_0-2}
  \begin{align}
    \Sigma_0 = & \exp \left( \left[ \bm \omega_{ref} \Delta t \right]_\times \right) \text{, } \label{eqn: Sigma_0}\\
    \Sigma_1 = &\text{ } \mathbf I \Delta t - \frac{\mathbf M_2}{\|\bm \omega_{ref}\|^2} \left( \Sigma_0 - \mathbf I - \mathbf M_2 \Delta t \right) \text{, } \label{eqn: Sigma_1} \\
    \Sigma_2 = &\text{ } \frac{1}{2} \mathbf I \Delta t^2 - \frac{1}{\|\bm \omega_{ref}\|^2} \left( \Sigma_0 - \mathbf I - \mathbf M_2 \Delta t - \frac{1}{2} \mathbf M_2^2 \Delta t^2 \right) \text{. }
  \end{align} 
\end{subequations}
The details of deriving \eqref{error-state time update} to (\ref{eqn:Sigma_0-2}) are given in Appendix A-B2.

\subsection{The SGW-GMM Radar Measurement Model} \label{sec:radar measurement model}

Let $\mathbf{Z}^r = \{ \mathbf{z}_{i}^r \}_{i=1}^{n^r}$ be the set of 3D position measurements from radar. Each measurement is generated from one of the $T$ reflectors based on probabilities $\bm \pi = \{ \pi_t \}_{t=1}^T$, with $\pi_t$ representing the probability associated with the $t$-th reflector. We only consider the case where there is a single object without generating any clutter measurements, and the radar measurement model can be expressed as:
\begin{equation}
  \mathbf z^r_{i} = \mathbf p + \mathbf w^r_{i} \text{, }
\end{equation}
where $\mathbf w^r_{i}$ is the measurement noise, and its condition probability density can be expressed as Gaussian mixture form:
\begin{equation}
  \label{radar noise model}
  p\left( \mathbf w^r_{i} \mid \bm \pi,\mathbf u, \mathbf x \right) = \sum_{t=1}^T \pi_t \mathcal N \left( \mathbf w^r_{i}; \mathbf R\{ \bm \theta \} \cdot \mathbf u_t, \mathbf Q \right) \text{, } 
\end{equation}
where $\mathbf Q$ denotes the radar measurement noise covariance.

\begin{figure}[tbp]
  \subfloat[Histograms of radar measurements]{
    \includegraphics[width=4cm]{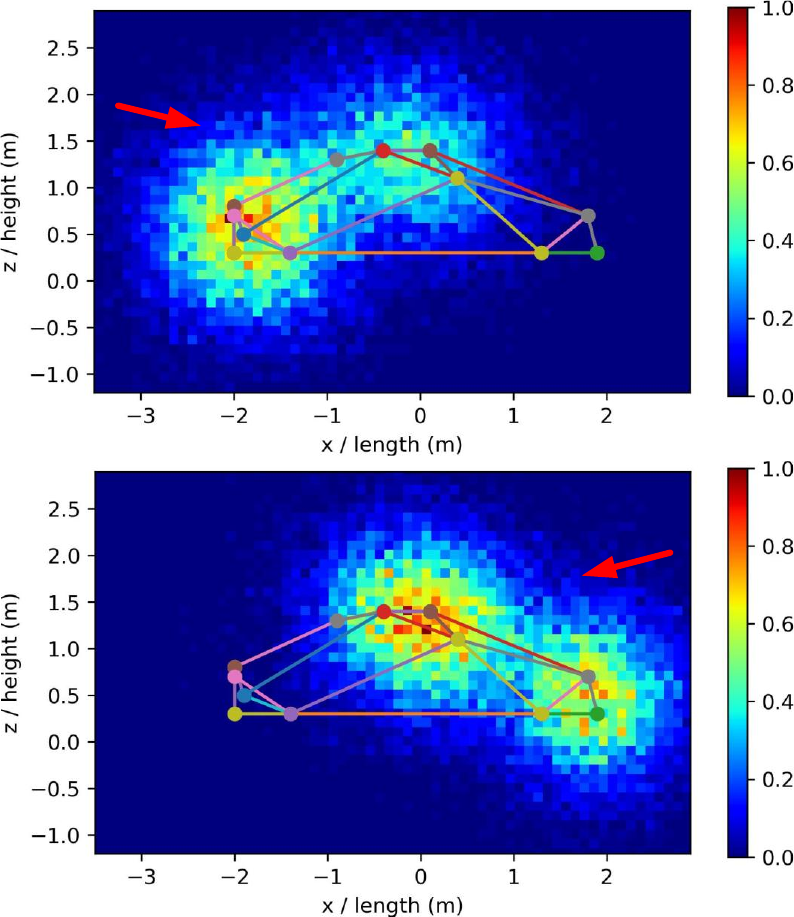}
  }
  \subfloat[SGW-GMM model]{
    \includegraphics[width=4cm]{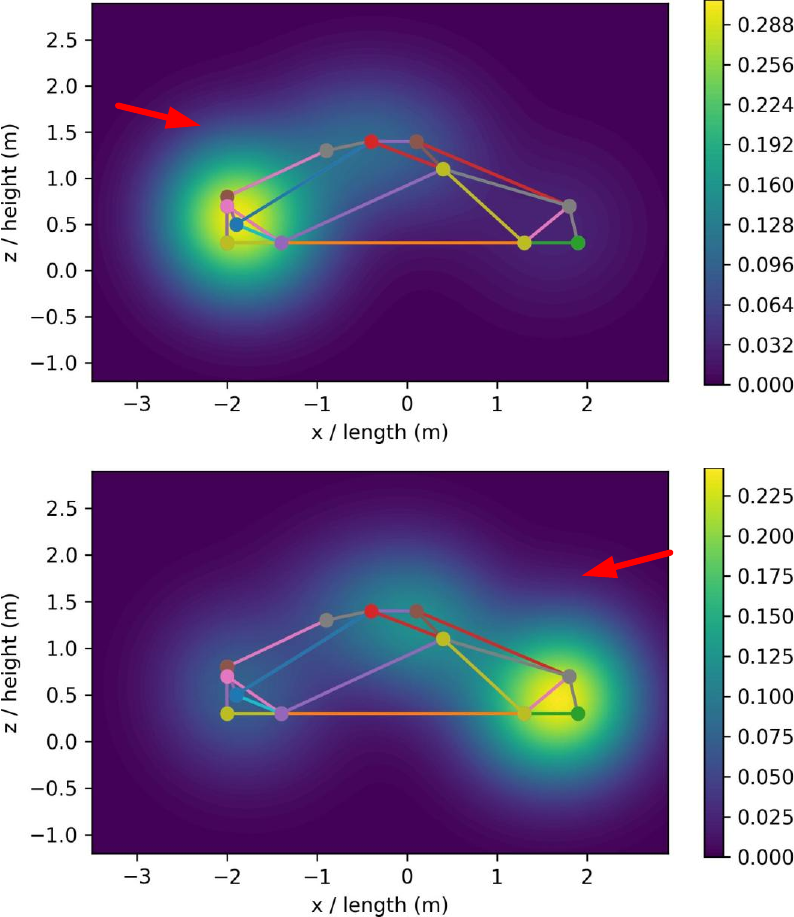}
    \label{subfig: SGW-GMM model}
  }
  \vspace{5pt}
  \caption{Histograms of radar measurements and probability density functions of the SGW-GMM radar measurement model $(\lambda = 1)$. For demonstration, the positions of skeleton knots and radar reflectors are assumed to be identical, denoted by knots connected with lines, while the observation directions are indicated by red arrows. It is important to note that our model does not constrain the positions of skeleton nodes and radar reflectors to be the same.}
  \label{fig:SGW-GMM measurement model}
\end{figure}

Different from how the weights in \eqref{radar noise model} are obtained in existing methods \cite{feldmann2010tracking, tuncer2022multi, scheel2018tracking, xia2021learning}, in this paper we propose a SGW-GMM measurement model that utilizes a spherical Gaussian function \cite{wang2009all} to weight the components, assigning higher weights to the components near the sensor direction and lower weights to those further away from the sensor direction:
\begin{equation}
  \label{pi_kt spherical gaussian}
  \pi_t \propto \exp\left( \lambda \times \left( -\frac{\mathbf p^\text{T} \cdot \left( \mathbf R\{ \bm \theta \} \cdot \mathbf u_t \right)}{\|\mathbf p\| \| \mathbf u_t \|} - 1 \right) \right) \text{, } 
\end{equation}
where the parameter $\lambda > 0$ represents the lobe sharpness, with a larger value of $\lambda$ indicating a more concentrated distribution of the described point cloud. For ease of calculation, we assume that $\pi_t$ is noiseless, and its value is calculated according to the vehicle's reference-state.

An illustration of the SGW-GMM measurement model for a car from rear-upper and front-upper observation directions is presented in Fig.~\ref{fig:SGW-GMM measurement model}. It is evident that the SGW-GMM model accurately captures the actual point cloud distribution.\footnote{Note that due to multipath effects in millimeter wave propagation, radar measurements may appear in positions where radar reflectors do not physically exist. Therefore, as shown in Fig.~\ref{subfig: SGW-GMM model}, the weights of negative side are non-zero, which means that occluded radar reflectors have relatively low probabilities of generating measurements, which helps describe radar measurements attributed to the multipath effect.} 
Furthermore, this model is also suitable for 3D EOT with sparse point clouds, because the vehicle's shape, represented by the positions of its skeleton knots, is partly supervised by the pixel keypoints detected through the vision algorithm, as demonstrated in Section \ref{sec:keypoints measurement model}. 


\subsection{The Pixel Keypoints Measurement Model} \label{sec:keypoints measurement model}

\begin{figure}[tbp]
  \centering
  \includegraphics[width=0.48\textwidth]{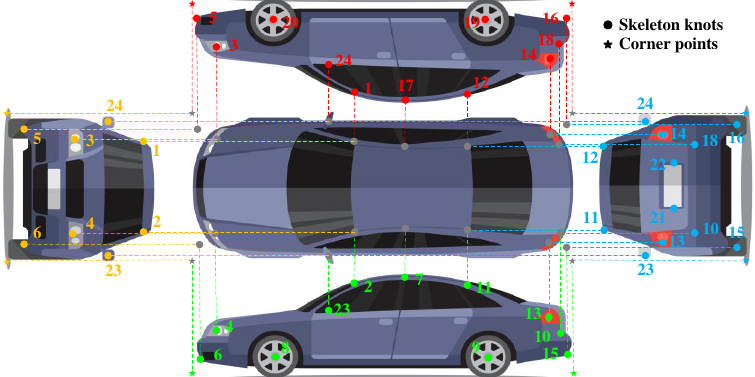}
  \vspace{5pt}
  \caption{Illustration of car's 24 skeleton knots and 4 corner points.}
  \label{fig:keypoints}
\end{figure}

Let $\mathbf{Z}^c = \{ \mathbf{z}_{i}^c \}_{i=1}^{n^c}$ be the set of 2D visual measurements obtained by the YOLOv8 pixel keypoints detection algorithm \cite{yolov8}. The measurements are categorized into two parts: measurements of the vehicle's skeleton knots $(\mathbf Z^{c, b})$ and measurements of the corner points of the vehicle's bottom $(\mathbf Z^{c, g} = \mathbf{Z}^c \setminus \mathbf Z^{c, b})$, as shown in Fig.~\ref{fig:keypoints}. These two parts are modeled separately:

\subsubsection{Modeling Skeleton Knot Measurements}

We denote the $i$-th pixel measurement of the vehicle's skeleton knots from the $s_i^b$-th component of the skeleton as $\mathbf z^{c, b}_i \in \mathbf Z^{c, b}$, which is modeled based on the vehicle's kinematic state and the position of the knot as follows:
\begin{equation}
  \label{keypoints measurement model 1}
  \mathbf z^{c, b}_i = h^{c, b} \left( \mathbf x, \bm \varpi_{s_i^b} \right) + \mathbf w^{c, b}_{s_i^b} \text{, } 
\end{equation}
where $\mathbf w^{c, b}_{s_i^b} \sim \mathcal N \left( 0, \mathbf Q_{s_i^b}^{c, b} \right)$ is the measurement noise. The nonlinear function $h^{c, b}()$ maps the coordinates of the knot $\bm \varpi_{s_i^b}$ from the VCS to the PCS using \eqref{eqn:rigid transformation} and \eqref{project transformation}, and the corresponding coordinates in the SCS and PCS are $\bm \varpi_{s_i^b}^{\mathcal{S}}$ and $\bm \varpi_{s_i^b}^{\mathcal{I}}$, respectively.

To linearize the model of skeleton knot measurements, we expand the function $h^{c, b}()$ around $\bm \varphi_{s_i^b} = \mathbb{E} \left[ \bm \varpi_{s_i^b} \right]$ and $\mathbf x = \mathbf x_{ref}$ using the Taylor series and neglect the quadratic or higher-order terms:
\begin{align}
  \label{base linearize model}
  \mathbf z^{c, b}_i \approx &\text{ } h^{c, b} \left( \mathbf x_{ref}, \bm \varphi_{s_i^b} \right) + \frac{\partial h^{c, b}}{\partial \delta \mathbf x} \bigg|_{\mathbf x = \mathbf x_{ref}, \bm \varpi_{s_i^b} = \bm \varphi_{s_i^b}} \delta \mathbf x \notag \\
  & + \frac{\partial h^{c, b}}{\partial \bm \varpi_{s_i^b}} \bigg|_{\mathbf x = \mathbf x_{ref}, \bm \varpi_{s_i^b} = \bm \varphi_{s_i^b}} \left( \bm \varpi_{s_i^b} - \bm \varphi_{s_i^b} \right) + \mathbf w^{c, b}_{s_i^b} \text{, } \notag \\
  = &\text{ } h^{c, b} \left( \mathbf x_{ref}, \bm \varphi_{s_i^b} \right) + \mathbf H^{c, b}_{\mathbf x} \delta \mathbf x \notag \\ 
  & + \mathbf H^{c, b}_{\bm \varpi_{s_i^b}} \left( \bm \varpi_{s_i^b} - \bm \varphi_{s_i^b} \right) + \mathbf w^{c, b}_{s_i^b} \text{, }
\end{align}
where
\begin{equation}
  \label{eqn:H_x^c,b}
  \mathbf H^{c, b}_{\mathbf x} = \frac{\partial \bm \varpi_{s_i^b}^{\mathcal{I}}}{\partial \bm \varpi_{s_i^b}^{\mathcal{S}}} \frac{\partial \bm \varpi_{s_i^b}^{\mathcal{S}}}{\partial \delta \mathbf x} \text{ , }
  \mathbf H^{c, b}_{\bm \varpi_{s_i^b}} = \frac{\partial \bm \varpi_{s_i^b}^{\mathcal{I}}}{\partial \bm \varpi_{s_i^b}^{\mathcal{S}}} \frac{\partial \bm \varpi_{s_i^b}^{\mathcal{S}}}{\partial \bm \varpi_{s_i^b}} \text{. }
\end{equation}
The common part of the two equations in \eqref{eqn:H_x^c,b} is acquired by taking the derivative of \eqref{project transformation}:
\begin{equation}
  \label{eqn:pixel Jacobian matrix}
  \frac{\partial \bm \varpi_{s_i^b}^{\mathcal{I}}}{\partial \bm \varpi_{s_i^b}^{\mathcal{S}}} = \begin{bmatrix} \frac{f/d_x}{\bm \varpi_{s_i^b}^{\mathcal{S}}[3]} & 0 & -\frac{\bm \varpi_{s_i^b}^{\mathcal{S}}[1] \cdot f/d_x}{\left(\bm \varpi_{s_i^b}^{\mathcal{S}}[3]\right)^2} \\ 0 & \frac{f/d_y}{\bm \varpi_{s_i^b}^{\mathcal{S}}[3]} & -\frac{\bm \varpi_{s_i^b}^{\mathcal{S}}[2] \cdot f/d_y}{\left(\bm \varpi_{s_i^b}^{\mathcal{S}}[3]\right)^2} \end{bmatrix} \text{, } 
\end{equation}
where $\bm \varpi_{s_i^b}^{\mathcal{S}}[i]$ represents the $i$-th element of the vector $\bm \varpi_{s_i^b}^{\mathcal{S}}$. According to the derivation of \eqref{eqn:rigid transformation}, we have that
\begin{equation}
  \frac{\partial \bm \varpi_{s_i^b}^{\mathcal{S}}}{\partial \delta \mathbf x} = \left[ \mathbf I_{3 \times 3}, \mathbf 0, \frac{\partial \left( \mathbf R\{\bm \theta\} \cdot \bm \varpi_t \right) }{\partial \delta \bm \theta}, \mathbf 0, \mathbf 0 \right] \text{ , } \frac{\partial \bm \varpi_{s_i^b}^{\mathcal{S}}}{\partial \bm \varpi_{s_i^b}} = \mathbf R\{\bm \theta\} \text{, }
\end{equation}
where, according to \cite[Eq. 188]{sola2017quaternion}, it holds that:
\begin{equation}
  \label{linearize R theta}
  \frac{\partial ( \mathbf R\{ \bm \theta \} \cdot \bm \varpi_{s_i^b} )}{\partial \delta \bm \theta} \bigg |_{\bm \theta=\bm \theta_{ref}} = -[\mathbf R \{ \bm \theta_{ref} \} \bm \varpi_{s_i^b}]_\times \mathbf J_l(\bm \theta_{ref}) \text{, }
\end{equation}
\begin{align}
  \mathbf J_l(\bm \theta_{ref}) = &\text{ } \mathbf I + \frac{1-\cos \| \bm \theta_{ref} \|}{\|\bm \theta_{ref} \|^2}[\bm \theta_{ref}]_\times  \notag \\
  & + \frac{\| \bm \theta_{ref} \| - \sin \| \bm \theta_{ref} \|}{\| \bm \theta_{ref} \|^3}[\bm \theta_{ref}]_\times^2 \text{. }
\end{align}

\subsubsection{Modeling Bottom Corner Point Measurements}

We denote the $i$-th pixel measurement of the vehicle's bottom corner points from the $s_i^g$-th corner point of the vehicle's bottom as $\mathbf z^{c, g}_i \in \mathbf Z^{c, g}$, which is modeled based on the vehicle's kinematic state as follows:
\begin{equation}
  \label{keypoints measurement model 2}
  \mathbf z^{c, g}_i = h_{s_i^g}^{c, g} \left( \mathbf x\right) + \mathbf w^{c, g}_{s_i^g} \text{, } 
\end{equation}
where $\mathbf w^{c, g}_{s_i^g} \sim \mathcal N \left( 0, \mathbf Q_{s_i^g}^{c, g} \right)$ is the corresponding noise, and $h_{s_i^g}^{c, g}()$ is a nonlinear function that maps the coordinates of vehicle's bottom from the VCS to the PCS.

Let the coordinates of the $s_i^g$-th corner points in the VCS be $\bm \xi_{s_i^g}^{\mathcal V} = \mathbf G_{s_i^g} \cdot \bm \xi$, where $\mathbf G_{s_i^g}$ represents the relative position of the corner point, given by:
\begin{equation}
  \mathbf G_{s_i^g} = \begin{bmatrix} \pm \frac{1}{2} & 0 \\ 0 & \pm \frac{1}{2} \\ 0 & 0 \end{bmatrix} \text{ , } s_i^g \in \{ 1, 2, 3, 4 \} \text{. }
\end{equation}
As a result, the output of function $h_{s_i^g}^{c, g} \left(\right)$ is $\bm \xi_{s_i^g}^{\mathcal I}$ by using \eqref{eqn:rigid transformation} and \eqref{project transformation}, and the coordinate in SCS is expressed as $\bm \xi_{s_i^g}^{\mathcal S}$.
Similar to \eqref{base linearize model}, for convenient calculation, we linearize \eqref{keypoints measurement model 2} using the first-order Taylor series approximation:
\begin{align}
  \label{bottom corner linearize model}
  \mathbf z^{c, g}_i & \approx h_{s_i^g}^{c, g} \left( \mathbf x_{ref} \right) + \frac{\partial h_{s_i^g}^{c, g}}{\partial \delta \mathbf x} \bigg|_{\mathbf x = \mathbf x_{ref}} \delta \mathbf x + \mathbf w^{c, g}_{s_i^g} \notag \\
  & = h_{s_i^g}^{c, g} \left( \mathbf x_{ref}\right) + \mathbf H^{c, g}_{\mathbf x, s_i^g} \delta \mathbf x + \mathbf w^{c, g}_{s_i^g} \text{, }
\end{align}
where
\begin{equation}
  \mathbf H^{c, g}_{\mathbf x, s_i^g} = \frac{\partial \bm \xi_{s_i^g}^{\mathcal{I}}}{\partial \bm \xi_{s_i^g}^{\mathcal{S}}} \frac{\partial \bm \xi_{s_i^g}^{\mathcal{S}}}{\partial \delta \mathbf x} \text{. }
\end{equation}
The first quotient of the above equation has the same form as \eqref{eqn:pixel Jacobian matrix}, while the second quotient is:
\begin{equation}
  \frac{\partial \bm \xi_{s_i^g}^{\mathcal{S}}}{\partial \delta \mathbf x} = \begin{bmatrix} \mathbf I_{3 \times 3} & \mathbf 0 & \frac{\partial ( \mathbf R\{\bm \theta\} \cdot \bm \xi_{s_i^g}^{\mathcal{V}} ) }{\partial \delta \bm \theta} & \mathbf 0 & \mathbf R\{\bm \theta\} \cdot \mathbf G_{s_i^g} \end{bmatrix} \text{. }
\end{equation}

\subsection{The Pseudo-measurement Models} \label{sec:pseudo-measurement model}

The pseudo-measurement models are introduced to process equality constraints between certain state variables \cite{julier2007kalman}. In this paper, we establish three constraints: the angular velocity constraint, the ground motion constraint, and the knot symmetry constraint.

\subsubsection{Angular Velocity Constraint}

Considering that the vehicle seldom involves roll rotation, which means that $\mathbf R\{ \bm \theta \} \cdot \mathbf u^d$ and $\bm \omega$ are approximately perpendicular. Therefore, the pseudo-measurement model can be expressed as:
\begin{equation}
  0 = \left( \mathbf R\{ \bm \theta \} \mathbf u^d \right)^\text{T} \bm \omega + w^{rot} \text{, }
\end{equation}
where $w^{rot} \sim \mathcal{N}\left( 0, \mathbf Q^{rot} \right)$. We linearize the above model according to \eqref{linearize R theta}, which yields
\begin{align}
  \label{linearized rotate constraint}
  0 = & \left( \mathbf R\{ \bm \theta_{ref} \} \mathbf u^d \right)^\text{T} \bm \omega_{ref} + \left( \mathbf R\{ \bm \theta_{ref} \} \mathbf u^d \right)^\text{T} \delta \bm \omega \notag \\
  & - \omega_{ref}^\text{T} \left[ \mathbf R\{ \bm \theta_{ref} \} \mathbf u^d \right]_\times \mathbf J_l\left( \bm \theta_{ref} \right) \delta \bm \theta + w^{rot} \text{. } 
\end{align}

\subsubsection{Ground Motion Constraint}


Without loss of generality, we assume that the corner points of the vehicle's bottom are on the ground, and the ground equation is represented as:
\begin{equation}
  \label{ground plane}
  \mathbf n^\text{T} \bm \beta + d = 0 \text{, }
\end{equation}
where $\mathbf n$ represents the unit normal vector of the ground plane, and $\bm \beta$ represents any point on the ground. Then the pseudo-measurement model of the $i$-th ($i \in \left[1, 4\right]$) corner is:
\begin{equation}
  0 = \mathbf n^\text{T} \left( \mathbf p + \mathbf R\{ \bm \theta \} \cdot \mathbf G_i \cdot \bm \xi \right) + d + w^{grnd} \text{, }
\end{equation}
where $w^{grnd} \sim \mathcal{N}\left( 0, \mathbf Q^{grnd} \right)$. Linearizing according to the \eqref{linearize R theta} gives:
\begin{align}
  \label{linearized bottom constraint}
  0 = &\text{ } \mathbf n^\text{T} \left( \mathbf p_{ref} + \mathbf R\{ \bm \theta_{ref} \} \cdot \mathbf G_i \cdot \bm \xi_{ref} \right) \notag \\
  & + \mathbf n^\text{T} \Big\{\delta \mathbf p + [\mathbf J_l(\bm \theta_{ref}) \delta \bm \theta]_\times \mathbf R \{ \bm \theta_{ref} \} \mathbf G_i \cdot \bm \xi_{ref} \notag \\
  & + \mathbf R\{ \bm \theta_{ref} \} \cdot \mathbf G_i \cdot \delta \bm \xi \Big\} + d + w^{grnd} \text{. }
\end{align}

\subsubsection{Knot Symmetry Constraint}

To infer the positions of the invisible knots from the visible knots, we leverage the left-right symmetric characteristic of the knots. Assuming the predefined symmetric knot for $ \bm \varpi_t$ is $\bm \varpi^{sym}_t$, which holds that $\left( \bm  \varpi^{sym}_t \right)^{sym}= \bm \varpi_t$, the pseudo-measurement model is:
\begin{equation}
  \bm \varpi^{sym}_t = \mathbf D \cdot \bm \varpi_t + \mathbf w^{sym}_t \text{, }
\end{equation}
where $\mathbf D = diag\left( 1, -1, 1 \right)$ is a diagonal matrix representing symmetric transformation, and $\mathbf w^{sym}_t \sim \mathcal{N} \left( 0, \mathbf Q^{sym} \right)$. 

\subsection{The ES Fusion Mechanism} \label{sec:ES model}


\begin{figure}[tbp]
  \centering
  \includegraphics[width=0.48\textwidth]{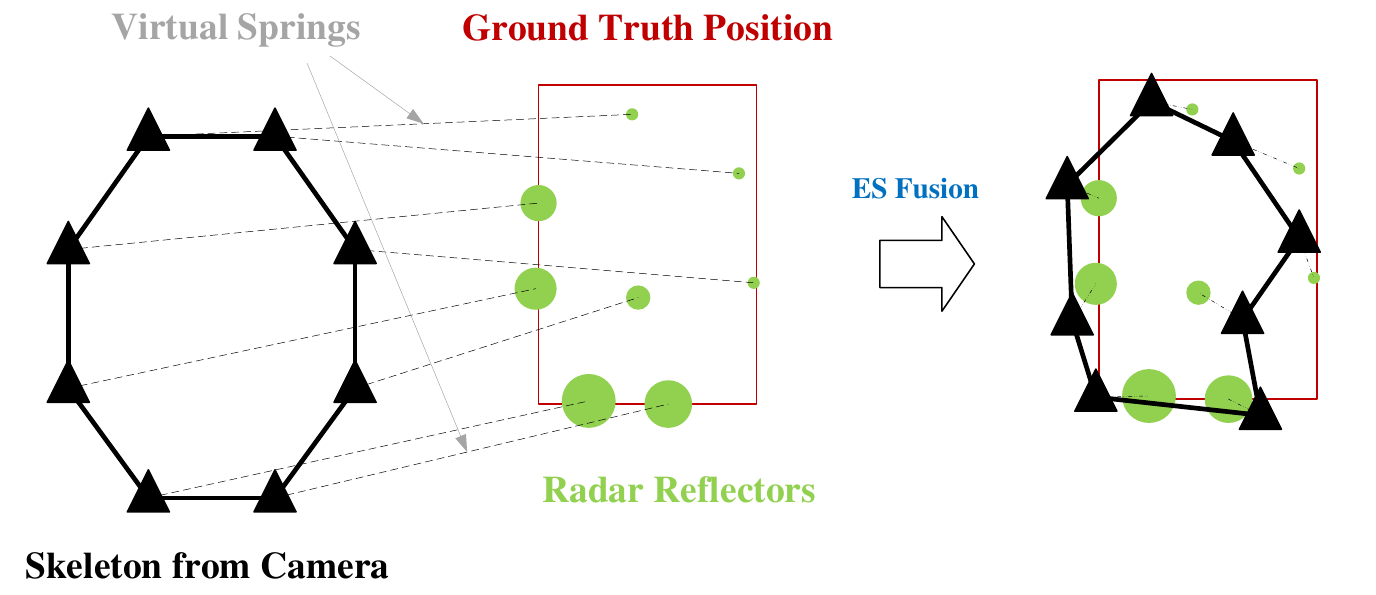}
  \vspace{5pt}
  \caption{Demonstration of the ES fusion mechanism. It is worth noting that while this mechanism enhances the accuracy of the vehicle's central position, it may also distort the shape of the vehicle's extent. However, during the updating stage, the extent is rectified by visual measurements.}
  \label{fig:ES_fusion}
\end{figure}

Based on the Bayesian framework, the measurement models update specific components of the object's state. For instance, the radar measurement model updates the coordinates of radar reflectors, while the pixel keypoints measurement model and the pseudo-measurement model of the ground plane update the coordinates of skeleton knots, along with the vehicle's length and width.

However, since radar and visual measurements are independent, the updated positions of radar reflectors lack correlation with the skeleton knots. While the positions of radar reflectors can effectively describe the object's position, they may not accurately represent its extent. Conversely, the vehicle skeleton accurately describes the object's extent but suffers from reduced position accuracy, particularly at longer distances.


To utilize the position information of radar reflectors, we introduce an ES fusion mechanism. This mechanism, as shown in Fig.~\ref{fig:ES_fusion}, leverages positions of radar reflectors and accurate extent information from pixel keypoints by constructing a spring-damping system that allows radar reflectors to pull the vehicle's extent to the correct position.

Assuming that there is a virtual spring connection with a natural length of 0 and an elastic coefficient of $\epsilon$ between each pair of reflectors and knots, the virtual mass of the knot is considered to be 1. To prevent system oscillation, damping with coefficient $\rho$ is introduced \cite{hallauer2016introduction}. Based on these assumptions, the skeleton knot is pulled by spring force from the reflector, and the dynamic model of the $t$-th component of the skeleton is described as follows:
\begin{equation}
  \label{ES dynamic model}
  \dot{\bm \vartheta}_t = \mathbf F_{\vartheta} \bm \vartheta_t + \mathbf w^{\vartheta} \text{, } 
\end{equation}
where
\begin{equation}
  \label{eqn: F_vartheta}
  \mathbf F_{\vartheta} = \begin{bmatrix}0 & 0 & 0 \\ 0 & 0 & \mathbf I \\ \epsilon \mathbf I & -\epsilon \mathbf I & -\rho \mathbf I \end{bmatrix} \text{ , } \mathbf w_{\vartheta} = \begin{bmatrix} \mathbf n_{u} \\ \mathbf 0 \\ \mathbf n_a \end{bmatrix} \text{, }
\end{equation}
with $\mathbf w_{\vartheta} \sim \mathcal N \left( 0, \mathbf W_{\vartheta} \right)$, and $\mathbf n_{u}$, $ \mathbf n_a$ denoting the position error of the reflector and the acceleration error of the knot, respectively.

We can observe that the form of \eqref{ES dynamic model} is similar to the transition model, described in \eqref{true kinematic model}. Hence, the influence of the ES fusion mechanism is reflected in the system's time update stage, and subsequently the time update for the vehicle's shape is obtained by integrating \eqref{ES dynamic model}:
\begin{subequations}
  \label{es model time update}
  \begin{align}
    \bm \vartheta_{k+1, t} & = \Phi_{\vartheta} \bm \vartheta_{k, t} + \mathbf w_{k}^{\vartheta} \text{, } \label{eqn: dynamic vartheta}\\
    \bm \vartheta_{0, t} & \sim \mathcal N \left( \hat{\bm \mu}_{0|0, t}, \hat{\bm \Sigma}_{0|0, t} \right) \text{, }
  \end{align} 
\end{subequations}
where $\mathbf w_{k}^{\vartheta} \sim \mathcal N \left( 0, \mathbf W_{\vartheta} \Delta t \right)$, the $\hat{\bm \mu}_{0|0, t}$ and $\hat{\bm \Sigma}_{0|0, t}$ are the initialized mean and covariance matrix of the $t$-th knot, respectively. The $\Phi_{\vartheta}$ is the transition matrix, with the closed form expression:
\begin{equation}
  \label{eqn:es transition matrix}
  \Phi_{\vartheta} =
  \begin{bmatrix}
    \mathbf I & \mathbf 0 & \mathbf 0 \\
    -\mathbf M_3 & \mathbf I + \mathbf M_3 & \mathbf M_4 \\
    \epsilon \mathbf M_4 & -\epsilon \mathbf M_4 & \mathbf I + \mathbf M_3 - \rho \mathbf M_4
  \end{bmatrix} \text{, } 
\end{equation}
where
\begin{subequations}
  \label{eqn: M_3 M_4}
  \begin{align}
    \mathbf M_3 = & \left\{ -1 + \exp(-\frac{\rho}{2 \Delta t}) \cosh \left( \frac{1}{2} \sqrt{-4\epsilon + \rho^2} \Delta t \right) \right. \notag \\
    & + \left. \frac{\exp(-\frac{\rho}{2 \Delta t}) \rho}{\sqrt{-4\epsilon + \rho^2}} \sinh \left( \frac{1}{2} \sqrt{-4\epsilon + \rho^2} \Delta t \right) \right\} \mathbf I \text{, } \\
    \mathbf M_4 = & \frac{2 \exp(-\frac{\rho}{2 \Delta t}) \sinh \left( \frac{1}{2} \sqrt{-4\epsilon + \rho^2} \Delta t \right)}{\sqrt{-4\epsilon+\rho^2}} \mathbf I \text{. }
  \end{align}
\end{subequations}
The details of deriving \eqref{es model time update} can be found in Appendix B.

\section{Inference via Variational Bayesian Approach}

We provide a conceptual overview before delving into the technical details. Specifically, the recursive Bayesian object state estimation relies on assumed density filtering, where in this work we assume that both the predicted and updated object's states follow Gaussian distributions. Here, the object's state at time $k-1$ is denoted as $\Theta_{k-1} = \{ \mathbf x_{k-1}, \bm \vartheta_{k-1}\}$, where $\bm \vartheta_{k-1} = \{ \bm \vartheta_{k-1, t} \}_{t=1}^T$.


We also let $\mathbf Z^{k-1} = \{ \mathbf Z^r_{1}, \mathbf Z^c_{1}, \dots, \mathbf Z^r_{k-1}, \mathbf Z^c_{k-1} \}$ represent the radar and visual measurements up to and including time step $k-1$. We assume that the posterior density at time $k-1$ is of the form:
\begin{align}
  \label{eqn:density form}
  p & \left( \Theta_{k-1} | \mathbf Z^{k-1} \right) \notag \\
  =&\text{ } \mathcal N \left( \mathbf x_{k-1}; \hat{\mathbf x}_{k-1|k-1}, \hat{\mathbf P}_{k-1|k-1} \right) \notag \\
  & \times \prod_{t=1}^T \mathcal{N}\left( \bm \vartheta_{k-1,t}; \hat{\bm \mu}_{k-1|k-1,t}, \hat{\bm \Sigma}_{k-1|k-1,t} \right) \text{. }
\end{align}

In this section, we show that the predicted density for the motion model introduced in Section \ref{sec:The 3D CTRV Motion Model} is also of the form \eqref{eqn:density form} without approximation. However, given the measurement model introduced in Section \ref{sec:radar measurement model} to \ref{sec:pseudo-measurement model}, the posterior density at time step $k$ has no closed-form solution. To enable recursive Bayesian filtering, we use variational inference to approximate the posterior such that the approximate density is also of the form \eqref{eqn:density form}. To simplify the derivation and present concise expressions, we introduce the following lemmas and symbols. For detailed proofs of these lemmas, please refer to Appendix C.

\begin{lemma}
  For a 3D vector $\mathbf{v}$ and a $3 \times n$ matrix $\mathbf{M}$, the following equation exists:
  \begin{equation}
    \left[ \mathbf v \right]_\times \cdot \mathbf M = -\left[ \mathbf M \right]_* diag_n \left( \mathbf v \right) \text{, }
  \end{equation}
  where the symbol $\left[ \cdot \right]_\ast$ represents the splicing for cross-product matrices for every column $(\mathbf M_{(i)}, i=1, \dots, n)$ of the matrix:
  \begin{equation}
    \left[ \mathbf M \right]_\ast = \left[ \begin{matrix} \left[\mathbf M_{(1)} \right]_\times & ,\dots,  & \left[\mathbf M_{(n)} \right]_\times \end{matrix} \right] \text{, }
  \end{equation}
  and the notation $diag_n(\cdot)$ represents the expansion of the vector $n$ times in a diagonal manner.
\end{lemma}

\begin{lemma}
  For two 3D vectors $\mathbf v$ and $\mathbf t$, and a $3 \times 3$ matrix $\mathbf M$, the following equation exists:
  \begin{equation}
    \left[ \mathbf v \right]_\times \cdot \mathbf M \cdot \mathbf t = -\left[ \mathbf M \right]_* \left[ \mathbf v \cdot \mathbf t^\text{T} \right]_{\star} \text{, }
  \end{equation}
  where the symbol $\left[ \cdot \right]_\star$ represents the recombination of the matrix into a column vector using row-major order.
\end{lemma}

\begin{lemma}
  For two 3D vectors $\mathbf v$ and $\mathbf t$, and a $3 \times 3$ symmetric positive definite (SPD) matrix $\mathbf M$, the following equation exists:
  \begin{align}
    \left[\mathbf v \right]_\times \cdot \mathbf M \cdot \left[\mathbf t \right]_\times^\text{T} = & \left[\mathbf v \right]_\times^\text{T} \cdot \mathbf M \cdot \left[\mathbf t \right]_\times \notag \\
    = & \left[ \mathbf L \right]_\ast diag_n \left( \mathbf v \cdot \mathbf t^\text{T} \right) \left[ \mathbf L \right]_\ast^\text{T} \text{, }
  \end{align}
  where the cholesky decomposition is $\mathbf M = \mathbf L \mathbf L^\text{T}$, and the $\mathbf L$ is also denoted as $\mathbf L = \sqrt{\mathbf M}$.
\end{lemma}

\subsection{Time Update} \label{sec:Time Update}

The time update for the kinematic state is divided into the reference-state time update and the error-state time update, as shown in \eqref{reference-state predict} and (\ref{error-state time update}):
\begin{align}
  \mathbf x_k & = \mathbf x_{ref,k} + \delta \mathbf x_{k} \notag \\
  & = f\left( \mathbf x_{ref,k-1} \right) + \Phi_{k-1} \delta \mathbf x_{k-1} + \mathbf w_{k-1} \text{. }
\end{align}
The resulting predicted density is:
\begin{equation}
  p\left( \mathbf x_k | \mathbf Z^{k-1} \right) = \mathcal N\left( \mathbf x_k; \hat{\mathbf x}_{k|k-1}, \hat{\mathbf P}_{k|k-1} \right) \text{, }
\end{equation}
where
\begin{subequations}
  \begin{align}
    \hat{\mathbf x}_{k|k-1} =& \text{ } f\left( \hat{\mathbf x}_{k-1|k-1} \right) \text{, }\\
    \hat{\mathbf P}_{k|k-1} =& \text{ } \Phi_k \hat{\mathbf P}_{k-1|k-1} \Phi_k^\text{T} + \mathbf W \Delta t \text{. }
  \end{align}
\end{subequations}
According to \eqref{es model time update}, the time update for the vehicle's shape is
\begin{equation}
  \bm \vartheta_{k, t} = \Phi_{\vartheta} \bm \vartheta_{k-1, t} + \mathbf w_{k-1}^{\vartheta} \text{. }
\end{equation}
The corresponding predicted density is then
\begin{equation}
  p\left( \bm \vartheta_{k, t} | \mathbf Z^{k-1} \right) = \mathcal N\left( \bm \vartheta_{k, t}; \hat{\bm \mu}_{k|k-1,t}, \hat{\bm \Sigma}_{k|k-1,t} \right) \text{. }
\end{equation}
where
\begin{subequations}
  \begin{align}
    \hat{\bm \mu}_{k|k-1,t} =& \text{ } \Phi_{\vartheta} \hat{\bm \mu}_{k-1|k-1,t} \text{, } \\
    \hat{\bm \Sigma}_{k|k-1,t} =& \text{ } \Phi_{\vartheta} \hat{\bm \Sigma}_{k-1|k-1,t} \Phi_{\vartheta}^\text{T} + \mathbf W_{\vartheta} \Delta t \text{. }
  \end{align}
\end{subequations}
Hence, the predicted joint probability density is:
\begin{align}
  \label{eqn:predict density}
  p & \left( \Theta_{k} | \mathbf Z^{k-1} \right) \notag \\
  =& \text{ }\mathcal N \left( \mathbf x_{k}; \hat{\mathbf x}_{k|k-1}, \hat{\mathbf P}_{k|k-1} \right) \notag \\
  &\times \prod_{t=1}^T \mathcal{N}\left( \bm \vartheta_{k,t}; \hat{\bm \mu}_{k|k-1,t}, \hat{\bm \Sigma}_{k|k-1,t} \right) \text{. }
\end{align}

\subsection{Likelihood Function}

The likelihood function consists of three parts corresponding to the three measurement models: the SGW-GMM radar measurement model, the keypoints measurement model, and the pseudo-measurement models.

\subsubsection{The Radar Likelihood Function}

The likelihood function of the radar measurements is obtained by multiplying the likelihood \eqref{radar noise model} of individual radar measurements:
\begin{align}
  \label{eqn: Ak 1}
  p(\mathbf Z_{k}^r|\Theta_{k}) &= \prod_{i=1}^{n^r_{k}}\sum_{t=1}^{T} \pi_{k,t}\mathcal N \left(\mathbf z_{k,i}^{r}; \bm \zeta_{k,t},\mathbf Q \right) \text{, }
\end{align}
where
\begin{subequations}
  \begin{align}
    \bm \zeta_{k,t} & = \mathbf p_{k}+ \mathbf R_k \mathbf u_{k, t} + [\mathbf J_{k} \delta \bm \theta_k]_\times \mathbf R_k \mathbf u_{k, t} \text{, } \\
    \mathbf R_k & = \mathbf R\{ \bm \theta_{ref, k} \} \text{, } \\
    \mathbf J_{k} & = \mathbf J_l(\bm \theta_{ref, k}) \text{. }
  \end{align}
\end{subequations}

Then we expand \eqref{eqn: Ak 1}, and introduce a latent variable $A_k = \{ \{ a_{it}\}_{i=1}^{n^r_k} \}_{t=1}^T$, which represent the data associations between measurements and radar reflectors, expressed as:
\begin{equation}
  a_{it}=\left\{\begin{array}{l l}{{1,}}&{{\mathrm{if~the~}t^{\text{th}} \text{ reflector generates } \mathbf z_{k,i}^r }}\\ {{0,}}&{{\mathrm{otherwise.}}}\end{array}\right.
\end{equation}
According to \cite{bishop2006pattern}, the equivalent expression for \eqref{eqn: Ak 1} is:
\begin{equation}
  \label{eqn: radar likelihood}
  p(\mathbf Z_{k}^r|\Theta_{k},A_{k})=\prod_{t=1}^{T}\prod_{i=1}^{n^r_{k}}\left[\mathcal N \left(\mathbf z_{k,i}^{r}; \bm \zeta_{k,t},\mathbf Q \right)\right]^{a_{it}} \text{. }
\end{equation}
The conditional probability density of $A_k$ is:
\begin{equation}
  \label{eqn: association likelihood}
  p(A_{k}|\Theta_{k})=\prod_{t=1}^{T}\prod_{i=1}^{n^r_{k}}(\pi_{k,t})^{a_{i t}} \text{, }
\end{equation}
where $\pi_{k,t}$ can be calculated according to \eqref{pi_kt spherical gaussian}, and normalization is required to ensure that $\sum_{t=1}^{T} \pi_{k,t} = 1$.

\subsubsection{The Keypoints Likelihood Function}

\begin{align}
  p(\mathbf Z_{k}^c|\Theta_{k}) = &\text{ } p(\mathbf Z_{k}^{c, b}|\Theta_{k}) \cdot p(\mathbf Z_{k}^{c, g}|\Theta_{k}) \notag \\
  = & \prod_{\mathbf z^{c, b}_{k, i} \in \mathbf Z_k^{c, b}} \mathcal N \left( \mathbf z_{k,i}^{c, b}; \mathbf h_{k,s_i^b}^{c, b}, \mathbf Q_{s_i^b}^{c, b} \right) \notag \\
  & \times \prod_{\mathbf z^{c, g}_{k, i} \in \mathbf Z_k^{c, g}} \mathcal N \left( \mathbf z_{k,i}^{c, g}; \mathbf h_{k,s_i^g}^{c, g}, \mathbf Q_{s_i^g}^{c, g} \right) \text{, }
\end{align}
where $\mathbf h_{k,s_i^b}^{c, b}$ and $\mathbf h_{k,s_i^g}^{c, g}$ are the noiseless parts of the right-hand sides of equations (\ref{base linearize model}) and (\ref{bottom corner linearize model}), respectively.

\subsubsection{The Pseudo-measurement Likelihood Function}

\begin{align}
  & p\left( z_k^{rot} | \Theta_k \right) p\left( \mathbf Z_k^{grnd} | \Theta_k \right) p\left( \mathbf Z_k^{sym} | \Theta_k \right)  \notag \\
  = & \text{ } \mathcal{N} \left( 0 ; h_k^{rot} , \mathbf Q^{rot} \right) \times \prod_{i=1}^4 \mathcal{N} \left( 0 ; h_{k, i}^{grnd, g} , \mathbf Q^{grnd} \right) \notag \\
  & \times \prod_{t=1}^T \mathcal{N} \left( \bm \varpi^{sym}_{k, t} ; \mathbf D \cdot \bm \varpi_{k, t} , \mathbf Q^{sym} \right) \text{, }
\end{align}
where $h_k^{rot}$ and $h_{k, i}^{grnd, g}$ are the noiseless parts of the right-hand sides of equations (\ref{linearized rotate constraint}) and (\ref{linearized bottom constraint}), respectively.

According to the above three likelihood functions, the joint likelihood function for all the measurements is:
\begin{align}
  \label{eqn:likelihood function}
  p(\mathbf Z_{k}|\Theta_{k},A_{k}) = &\text{ } p(\mathbf Z_{k}^r, \mathbf Z_k^c, z_k^{rot} |\Theta_{k},A_{k}) \notag \\
  = &\text{ } p(\mathbf Z_{k}^r|\Theta_{k},A_{k}) p(\mathbf Z_k^c|\Theta_{k}) p\left( z_k^{rot} | \Theta_k \right) \notag \\
  & \times p\left( \mathbf Z_k^{grnd} | \Theta_k \right) p\left( \mathbf Z_k^{sym} | \Theta_k \right) \text{. }
\end{align}

\subsection{Measurement Update}

Given the predicted density \eqref{eqn:predict density} and likelihood functions \eqref{eqn:likelihood function}, the joint posterior probability density of the object's state can be computed using the Bayes rule
\begin{align}
  \label{eqn: post prob}
  p(\Theta_k, A_k| \mathbf Z^{k}) & = \frac{p(\Theta_k, A_k, \mathbf Z_k| \mathbf Z^{k-1})}{p(\mathbf Z_k|\mathbf Z^{k-1})} \notag \\
  & = \frac{p(\Theta_k|\mathbf Z^{k-1}) p(A_k| \Theta_k) p(\mathbf Z_k| \Theta_k, A_k)}{p(\mathbf Z_k|\mathbf Z^{k-1})} \text{. }
\end{align}

The above posterior probability density has no analytical expression, so we adopt a variational Bayesian approach \cite{orguner2012variational} to find a tractable approximation such that the approximated density $p\left( \Theta_k | \mathbf Z^{k} \right)$ is still of the form \eqref{eqn:density form}, which is required for recursive Bayesian filtering. The posterior probability density can be approximated as the product of three marginal densities according to mean-field theory:
\begin{equation}
  \label{eqn:mean-field theory}
  p(\Theta_k, A_k| \mathbf Z^{k}) \approx q_A(A_k) q_{\mathbf x}(\mathbf x_k) q_{\bm \vartheta}(\bm \vartheta_k) \text{. }
\end{equation}

A nice property of the factorized form \eqref{eqn:mean-field theory} is that the posterior density $p(\Theta_k| \mathbf Z^{k})$ can be easily obtained from \eqref{eqn:mean-field theory} by marginalizing out the data association variables $A_k$. In addition, the approximate distribution can be derived by minimizing the KL divergence \cite{csiszar1975divergence} between the approximate distribution and the true posterior distribution
\begin{equation}
  \argmin_{q_A, q_{\mathbf x}, q_{\bm \vartheta}} \mathrm{KL} \left(q_A\left( A_k \right) q_{\mathbf x}\left( \mathbf x_k \right) q_{\bm \vartheta}\left( \bm \vartheta_k \right) \| p\left( \Theta_k, A_k| \mathbf Z^{k} \right) \right) \text{. }
\end{equation}
The optimal solution of the above problem satisfies:
\begin{equation}
  \label{optimal solution}
  \log q_{\varkappa}(\kappa_{k})=\mathbb{E}_{\\ \varkappa^*}\!\left[\log p(\Theta_k, A_k, \mathbf Z_k| \mathbf Z^{k-1})\right]+c_{\varkappa^*} \text{, }
\end{equation}
where $\varkappa \in \{ A, \mathbf x, \bm \vartheta \}$, $\varkappa^*$ represents a set of all elements excluding $\varkappa$, and $c_{\varkappa^*}$ represents a constant unrelated to $\varkappa$. For example, $\mathbb{E}_{A^*}[\cdot]$ represents the expectation with respect to $\mathbf x$ and $\bm \vartheta$. 

We use iterative coordinate ascent to solve the optimization problem \eqref{optimal solution} for each probability distribution $q_A$, $q_{\mathbf x}$, $q_{\bm \vartheta}$ in turn. We assume that the probability density function for the variables $\mathbf x$ and $\bm \vartheta$ in the $\iota$-th iteration are
\begin{subequations}
  \label{eqn: final state prob}
  \begin{align}
    q^{(\iota)}_{\mathbf x}(\mathbf x_k) & = \mathcal N \left( \mathbf x_{k}; \hat{\mathbf x}_{k|k}^{(\iota)}, \hat{\mathbf P}_{k|k}^{(\iota)} \right) \text{, } \\
    q^{(\iota)}_{\bm \vartheta}(\bm \vartheta_k) & = \prod_{t=1}^T \mathcal{N}\left( \bm \vartheta_{k,t}; \hat{\bm \mu}_{k|k,t}^{(\iota)}, \hat{\bm \Sigma}_{k|k,t}^{(\iota)}  \right) \text{. }
  \end{align}
\end{subequations}
Notably, we initialize \eqref{eqn: final state prob} using predicted densities, where $q^{(0)}_{\mathbf x}(\mathbf x_k) = p\left( \mathbf x_k | \mathbf Z^{k-1} \right)$ and $q^{(0)}_{\bm \vartheta}(\bm \vartheta_k) = p\left( \bm \vartheta_{k, t} | \mathbf Z^{k-1} \right)$. Subsequently, we proceed to describe how to calculate the $(\iota+1)$-th iteration. For notational brevity, we use the shorthand notation $\left(\mathbf v\right)^\text{T} \mathbf Q^{-1} (\cdot)$ notation for $\left(\mathbf v\right)^\text{T} \mathbf Q^{-1} \left(\mathbf v\right)$.


\subsubsection{Computing $q^{(\iota+1)}_{A}(A_{k})$}
By substituting $A$ for $\varkappa$ in \eqref{optimal solution}, we obtain
\begin{align}
  &\text{ } \log q^{(\iota+1)}_{A}(A_{k}) = \mathbb{E}_{A^*}\left[\log p(\Theta_k, A_k, \mathbf Z_k| \mathbf Z^{k-1})\right]+c_{A^*} \notag \\
  = & \sum_{t=1}^{T} \sum_{i=1}^{n^r_k} a_{it} \left( \log \pi_{k,t}^{(\iota)} - \frac{1}{2} \mathbb{E} \left[ \left( \mathbf z_{k,i}^r - \bm \zeta_{k,t}^{(\iota)} \right)^\text{T} \mathbf Q^{-1} (\cdot) \right] \right) + c_{A^*} \text{. }
\end{align}
Then, applying the exponential to both sides, we can obtain the $(\iota + 1)$-th iteration of $q^{(\iota+1)}_{A}(A_{k})$:
\begin{equation}
  \label{eqn: update association}
  q^{(\iota+1)}_{A}(A_{k}) = \prod_{t=1}^T \prod_{i=1}^{n^r_k} \left(\upsilon_{it}^{(\iota)}\right)^{a_{it}} \text{, }
\end{equation}
with
\begin{align}
  \upsilon_{it}^{(\iota)} &= \frac{\rho_{it}^{(\iota)}}{\sum_{t=1}^\text{T} \rho_{it}^{(\iota)}} \text{, } \\
  \label{eqn:rho in q_A}
  \rho_{it}^{(\iota)} &= \exp \left\{ \log \pi_{k,t}^{(\iota)} - \frac{1}{2} \mathbb{E} \left[ \left( \mathbf z_{k,i}^r - \bm \zeta_{k,t}^{(\iota)} \right)^\text{T} \mathbf Q^{-1} (\cdot) \right] \right\} \text{. } 
\end{align}
The expression of the expectation in \eqref{eqn:rho in q_A} can be simplified using Lemmas 1 to 3, yielding:
\begin{align}
  & \mathbb{E} \left[ \left( \mathbf z_{k,i}^r - \bm \zeta_{k,t}^{(\iota)} \right)^\text{T} \mathbf Q^{-1} (\cdot) \right] \notag \\
  = & \left( \mathbf z_{k,i}^r - \mathbb{E} \left[ \bm \zeta_{k,t}^{(\iota)} \right] \right)^\text{T}  \mathbf Q^{-1} (\cdot) + \mathrm{tr} \left\{\mathbf Q^{-1} \mathrm{Cov}\left[ \mathbf p_k^{(\iota)} \mathbf p_k^{(\iota),\text{T}} \right]  \right\} \notag \\
  &+ \mathrm{tr} \left\{ \mathbf R^{(\iota),\text{T}}_k  \mathbf Q^{-1} \mathbf R^{(\iota)}_k \mathrm{Cov}\left[ \mathbf u^{(\iota)}_{k,t} \mathbf u^{(\iota),\text{T}}_{k,t} \right] \right. \notag  \\
  &+ \overline{\mathbf R}^{(\iota),\text{T}}_k \mathrm{diag}_3\left( \mathbf J_{k}^{(\iota)} \mathbb{E} \left[ \delta \bm \theta_k^{(\iota)} \delta \bm \theta_k^{(\iota),\text{T}} \right] \mathbf J_{k}^{(\iota),\text{T}} \right) (\cdot) \mathbb{E} \left[ \mathbf u^{(\iota)}_{k,t} \mathbf u^{(\iota),\text{T}}_{k,t} \right] \notag \\
  &- \left. 2 \mathbf Q^{-1} \left[ \mathbf R^{(\iota)}_k \mathbb{E} \left[ \mathbf u^{(\iota)}_{k,t} \right] \right]_\times \mathbf J_{k}^{(\iota)} \mathbb{E} \left[ \delta \bm \theta_k^{(\iota)} \delta \mathbf p_k^{(\iota),\text{T}} \right] \right\} \text{, }
\end{align}
where we denote
\begin{subequations}
  \begin{align}
    \mathbf R^{(\iota)}_k & = \mathbf R\{ \bm \theta_{ref, k}^{(\iota)} \} \text{, } \\
    \overline{\mathbf R}^{(\iota)}_k & = \left[ \sqrt{ \mathbf Q^{-1} } \right]_\ast^\text{T} \mathbf R^{(\iota)}_k \text{, } \\
    \mathbf J_{k}^{(\iota)} & = \mathbf J_l(\bm \theta_{ref, k}^{(\iota)}) \text{. }
  \end{align}
\end{subequations}

\subsubsection{Computing $\mathbb{E}_{A}\left[\log p(\Theta_k, A_k, \mathbf Z_k| \mathbf Z^{k-1})\right]$}
The expectation of the log joint probability density with respect to variable $A$ for the $(\iota+1)$-th iteration is
\begin{align}
  \label{eqn:E_A} 
  & \mathbb{E}_{A}\left[\log p(\Theta_k, A_k, \mathbf Z_k| \mathbf Z^{k-1})\right] \notag  \\
  =& - \frac{1}{2} \Bigg\{ \sum_{t=1}^T \bigg\{ \mathrm{tr} \left( \overline{\mathbf Z}^{(\iota+1)}_{k,t} \mathbf Q^{-1} \right) \notag \\
  &+ \left( \overline{\mathbf z}^{(\iota+1)}_{k,t} - \bm \zeta_{k,t} \right)^\text{T} \left( \frac{\mathbf Q}{n^{(\iota+1)}_{k,t}} \right)^{-1} (\cdot) \notag \\
  &+ \left( \bm \vartheta_{k,t}-\hat{\bm \mu}^{(\iota)}_{k|k,t} \right)^\text{T} \left( \hat{\bm \Sigma}^{(\iota)}_{k|k,t} \right)^{-1} (\cdot) \notag \\
  &+ \left( \bm \varpi_{k,t}^{sym} - \mathbf D \cdot \bm \varpi_{k,t} \right)^\text{T} \left( \mathbf Q^{sym} \right)^{-1} (\cdot) \bigg\} \notag \\
  &+ \left( \mathbf x_k - \hat{\mathbf x}^{(\iota)}_{k|k} \right)^\text{T} \left( \hat{\mathbf P}^{(\iota)}_{k|k} \right)^{-1} (\cdot) \notag \\
  &+ \sum_{\mathbf z^{c, b}_{k, i} \in \mathbf Z_k^{c, b}} \left( \mathbf z^{c, b}_{k, i} - \mathbf h_{k,s_i^b}^{c, b, (\iota)} \right)^\text{T} \left( \mathbf Q_{s_i^b}^{c, b, (\iota)} \right)^{-1} (\cdot) \notag \\
  &+ \sum_{\mathbf z^{c, g}_{k, i} \in \mathbf Z_k^{c, g}} \left( \mathbf z_{k,i}^{c, g} - \mathbf h_{k,s_i^g}^{c, g, (\iota)} \right)^\text{T} \left( \mathbf Q_{s_i^g}^{c, g, (\iota)} \right)^{-1} (\cdot) \notag \\
  &+ \left( h_k^{rot, (\iota)} \right)^2 \left( \mathbf Q^{rot} \right)^{-1} + \sum_{i=1}^{4} \left( h_{k,i}^{grnd, g, (\iota)} \right)^2 \left( \mathbf Q^{grnd} \right)^{-1} \Bigg\} \notag \\
  &+ c_{\mathbf x^*} + c_{\bm \vartheta^*} \text{, } 
\end{align}
where
\begin{subequations}
  \label{eqn:n_kt_Z_line}
  \begin{align}
    n^{(\iota+1)}_{k,t} & =\sum_{i=1}^{n^r_{k}}\upsilon^{(\iota)}_{i t} \label{eqn:n_kt} \text{, } \\
    \overline{\mathbf z}^{(\iota+1)}_{k,t} & = \frac{1}{n^{(\iota+1)}_{k,t}}\sum_{i=1}^{n^r_{k}} \upsilon^{(\iota)}_{i t} \mathbf z_{k,i}^{r} \text{, } \\
    \overline{\mathbf Z}^{(\iota+1)}_{k,t} & = \sum_{i=1}^{n^r_k} \upsilon^{(\iota)}_{i t} \left( \mathbf z_{k,i}^r - \overline{\mathbf z}^{(\iota+1)}_{k,t} \right) \left( \mathbf z_{k,i}^r - \overline{\mathbf z}^{(\iota+1)}_{k,t} \right)^\text{T} \text{. } \label{eqn:Z_line_kt}
  \end{align}
\end{subequations}
For the detailed simplification of \eqref{eqn:E_A}-\eqref{eqn:n_kt_Z_line}, please refer to Appendix D.

\subsubsection{Computing $q^{(\iota + 1)}_{\mathbf x}(\mathbf x_k)$ and $q^{(\iota + 1)}_{\bm \vartheta}(\bm \vartheta_k)$}
By replacing $\varkappa$ in \eqref{optimal solution} with $\mathbf x$ and $\bm \vartheta$ respectively, we obtain
\begin{subequations}
  \begin{align}
    \log q^{(\iota+1)}_{\mathbf x}(\mathbf x_k) = &\text{ } \mathbb{E}_{\bm \vartheta}\left[ \mathbb{E}_{A}\left[\log p(\Theta_k, A_k, \mathbf Z_k| \mathbf Z^{k-1})\right] \right] + c_{\mathbf x^*} \text{, } \\
    \log q^{(\iota+1)}_{\bm \vartheta}(\bm \vartheta_k) = &\text{ } \mathbb{E}_{\mathbf x}\left[ \mathbb{E}_{A}\left[\log p(\Theta_k, A_k, \mathbf Z_k| \mathbf Z^{k-1})\right] \right] + c_{\bm \vartheta^*} \text{. }
  \end{align}
\end{subequations}
After applying exponential to both sides, we represent the closed-form expressions of the probability densities as weighted combinations of prior information and measurements. The detailed expressions can be found Appendix E. Here, we offer simplified versions
\begin{subequations}
  \label{eqn: update_state}
  \begin{align}
    q^{(\iota+1)}_{\mathbf x}(\mathbf x_k) = &\text{ } \mathcal N \left( \mathbf x_{k}; \hat{\mathbf x}_{k|k}^{(\iota+1)}, \hat{\mathbf P}_{k|k}^{(\iota+1)} \right) \text{, } \label{eqn: update_x} \\
    q^{(\iota+1)}_{\bm \vartheta}(\bm \vartheta_k) = & \prod_{t=1}^T \mathcal {N} \left( \bm \vartheta_{k,t}; \hat{\bm \mu}_{k|k,t}^{(\iota+1)}, \hat{\bm \Sigma}_{k|k,t}^{(\iota+1)} \right) \text{. } \label{eqn: update_vartheta}
  \end{align}
\end{subequations}

Finally, after $N_{vb}$ rounds of iterations, the posterior probability density of the object's state can be written as:
\begin{equation}
  p(\Theta_k| \mathbf Z^{k}) \approx q^{(N_{vb})}_{\mathbf x}(\mathbf x_k) \cdot q^{(N_{vb})}_{\bm \vartheta}(\bm \vartheta_k) \text{. }
\end{equation}
The parameters of the probability density function represent the inferred mean and covariance of the object's state.

\section{Simulation and Experimental Results}

\subsection{Simulation Results}

The performance of the proposed algorithm is evaluated in several simulation scenarios using the CARLA simulator \cite{dosovitskiy2017carla}, against three other 3D EOT methods using only radar measurements\footnote{Python implementation of all the simulation methods and data are available at https://github.com/RadarCameraFusionTeam-BUPT/ES-EOT}: GPEOT and GPEOT-P as described in \cite{kumru2021three}, and the ellipsoid-based method MEM-EKF* \cite{yang2019tracking}. In this section, we refer to our proposed method as ``ES-EOT'' (short for elastic skeleton extended object tracker). Additionally, to highlight the effectiveness of our proposed 3D CTRV motion model and the ES fusion mechanism, we conducted ablation studies in the same test scenarios. Specifically, by setting $\bm\omega_{ref,k}$ in \eqref{eqn:state predict1} and (\ref{eqn:Sigma_0-2}) to 0 while maintaining consistency in other aspects, a constant velocity (CV) motion model is constructed as a comparison method. In addition, we also compared the situation without using the ES fusion mechanism.


We conducted roadside experiments using radar and camera sensors, analyzing two types of trajectories involving two different vehicle types: a bus changing lanes and a sedan car turning around. Trajectories are automatically generated by the simulator using specified start and end points, as well as intermediate points along the path. For vision keypoints detections, we first annotate the true 3D keypoints (24 knots and 4 bottom corner points) on vehicle models used within the simulator, and then trained the YOLOv8 network\footnote{The architecture of the network used in this paper is the same as \cite{yolov8}, which has image input and keypoints outputs. However, the key distinction lies in training the network on our dedicated Carla dataset.} to predict the pixel coordinates of the visible keypoints. For radar measurements, we synthesize radar points with 3D positions derived from visible skeleton knots, and introduce a measurement noise with covariance matrix of $0.5\mathbf I_{3\times3}$. In order to evaluate the algorithm's performance across varying radar point densities, we introduce different densities by assuming a Poisson distribution for the number of radar points in each frame. The Poisson rates $(\alpha)$ are set at 1, 5, and 10, respectively, while ensuring a minimum of one radar point. These experiments are repeated 50 times for each scenario and Poisson rates setting.

We evaluated the algorithm's performance by employing the Intersection-Over-Union (IOU) metric \cite{rezatofighi2019generalized} of the vehicle's shape projections onto the VCS's ``xy-plane'', ``xz-plane'', and ``yz-plane'':
\begin{equation}
  \text{IOU}_{\text{*-plane}}\left( V, \hat{V} \right) = \frac{\text{area}\left( \text{proj}_{\text{*}} \left( V \right) \cap \text{proj}_{\text{*}} \left( \hat{V} \right) \right)}{\text{area}\left( \text{proj}_{\text{*}} \left( V \right) \cup \text{proj}_{\text{*}} \left( \hat{V} \right)\right)} \text{, }
\end{equation}
where $V$ represents the true 3D shape of the vehicle, and $\hat{V}$ is the estimated shape. The function $\text{proj}_{\text{*}} (\cdot)$ projects the 3D shape onto the ``*-plane'', and the function $\text{area} (\cdot)$ calculates the area of the resulting 2D shape. The IOU metric reflects the accuracy of both the vehicle's shape and kinematic state estimation, with a higher IOU indicating better estimates.

We calculate the average performance in each tracking experiment, which includes the mean value of the IOU and the root mean squared error (RMSE) of the velocity:
\begin{align}
  \overline{\text{IOU}}_{\text{*-plane}} = &\text{ } \frac{1}{N} \sum_{i=1}^N \text{IOU}_{\text{*-plane}}\left( V_i, \hat{V}_i \right) \text{, } \\
  \text{RMSE}_v = & \sqrt{\frac{1}{N}\sum_{i=1}^N \left( v_i - \hat{v}_i \right)^2} \text{, }
\end{align}
where $v_i$ and $V_i$ represent the true velocity and shape of the vehicle in the $i$-th frame, while $\hat{v}_i$ and $\hat{V}_i$ correspond to the estimated velocity and shape, respectively. The $N$ is the total number of frames.

We set the parameters of process noise described in \eqref{error-state kinematic} and \eqref{ES dynamic model} as $\mathbf W = \mathrm{diag}\left( \mathbf 0_{3 \times 3}, 0.01, \mathbf 0_{3 \times 3}, \mathbf I_{3 \times 3}, 0.1 \mathbf I_{2 \times 2} \right)$ and $\mathbf W_{\vartheta} = \mathbf I_{9 \times 9}$, respectively. The parameters of ES fusion mechanism described in Section \ref{sec:ES model} are $\epsilon = 100$ and $\rho = 20$, ensuring critical damping behavior. The hyper-parameters of the algorithm are set to $\mathbf Q = 0.5 \mathbf I_{3 \times 3}$, $\mathbf Q_{s_i^b}^{c, b} = \mathbf Q_{s_i^g}^{c, g} = 5\mathbf I_{2 \times 2}$, $\mathbf Q^{rot} = 0.1$, $\mathbf Q^{grnd} = 10^{-4}$, $\mathbf Q^{sym} = 0.05 \mathbf I_{3 \times 3}$, $\mathbf D = \mathrm{diag} \left( 1, -1, 1 \right)$, $\mathbf u_d = \left[ 1, 0, 0 \right]$. The ground equation described in \eqref{ground plane} is obtained using our previous FusionCalib method \cite{deng2023fusioncalib}. The vehicle's shape has $T = 24$ components, and we perform $N_{vb} = 3$ iterations for variational Bayesian inference.

\subsubsection{Scenario of Changing Lane}

\begin{figure}[tbp]
  \begin{minipage}[t]{0.5\textwidth}
    \centering
    \subfloat[]{
      \includegraphics[width=4.125cm,height=2.5cm]{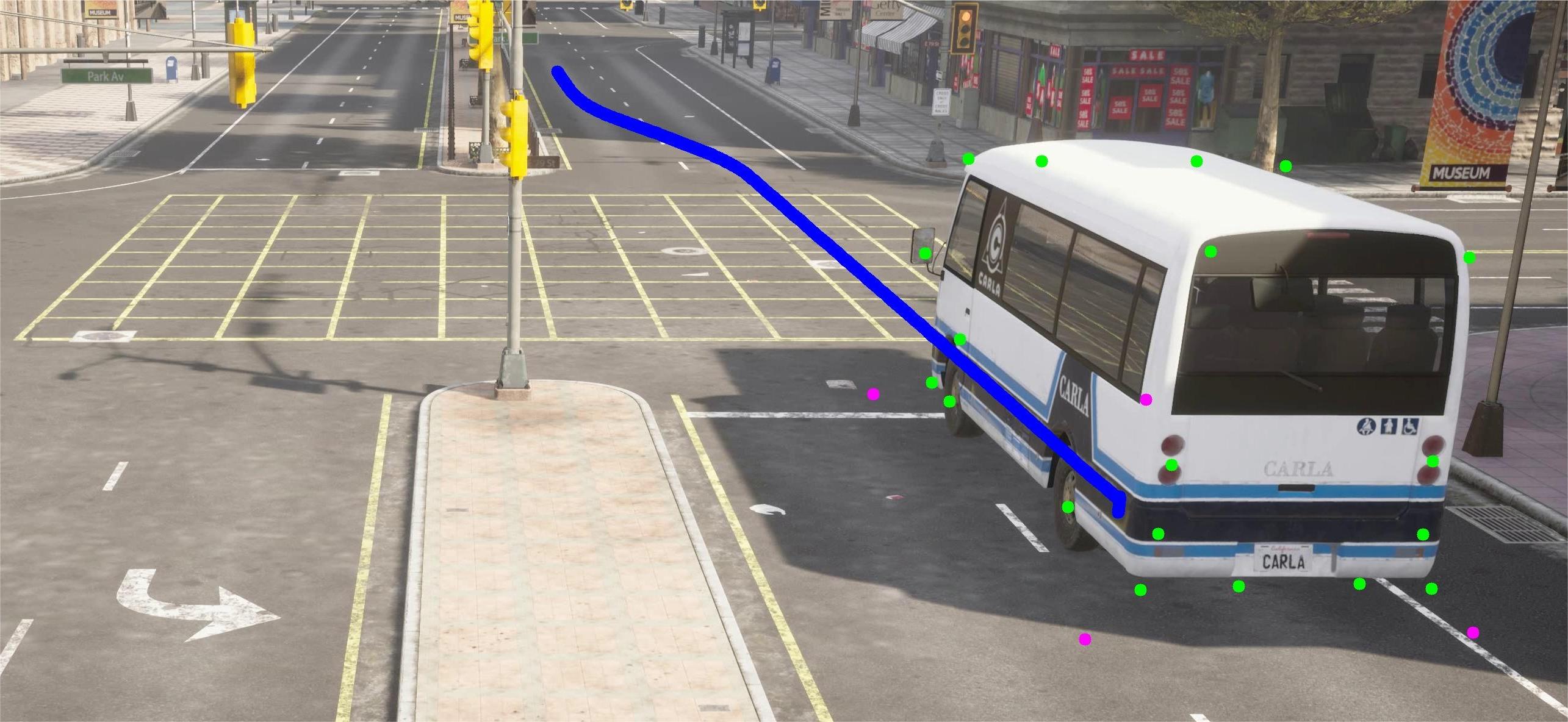}
    }
    \subfloat[]{
      \includegraphics[width=4.125cm,height=2.5cm]{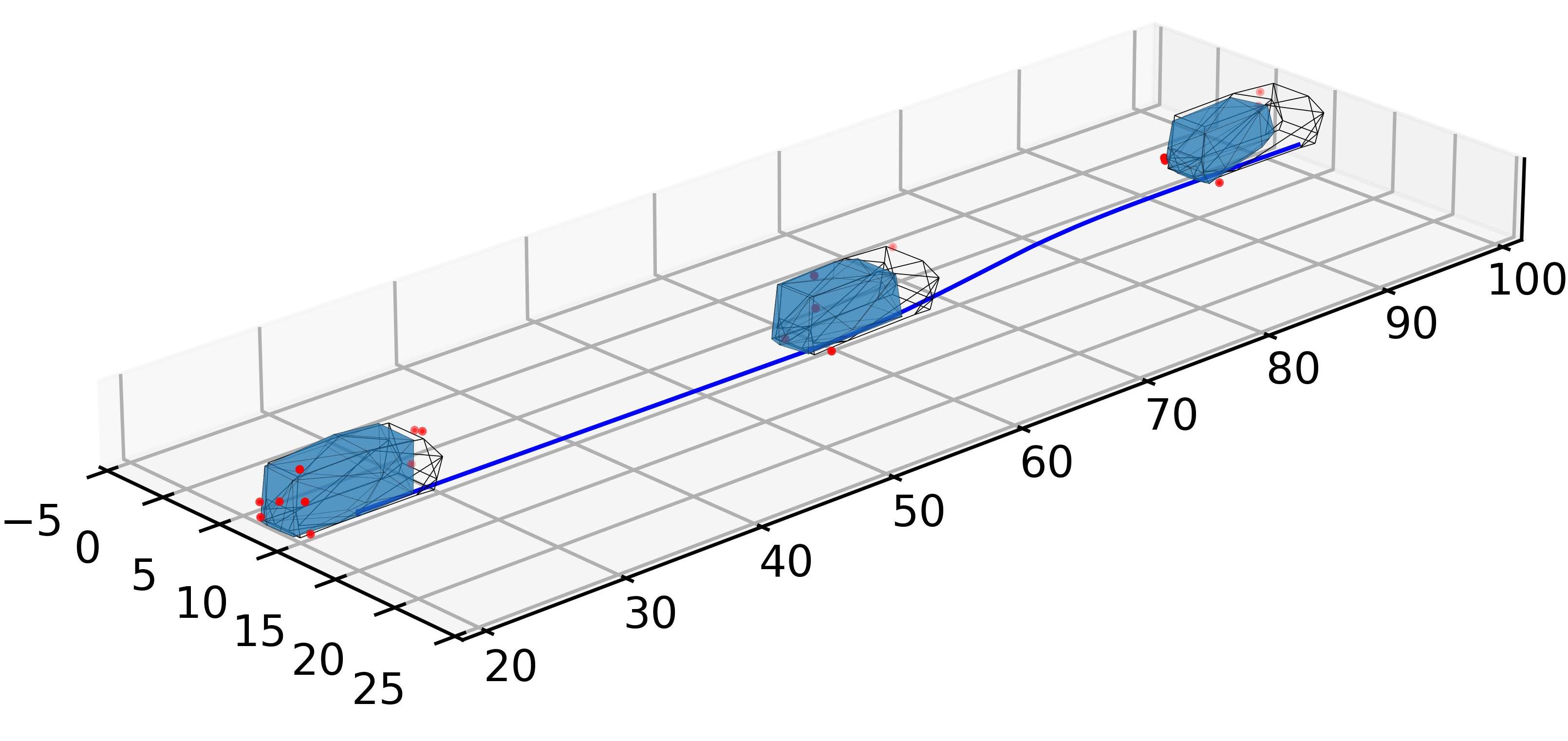}
    }
    \caption{The detections and results from the changing lane scenario. In Fig. (a), the blue line is the actual vehicle trajectory, green points indicate detected vehicle skeleton knots, and magenta points represent detected bottom corner points. Fig. (b) shows radar detections $(\alpha = 10)$ as red points, the vehicle shape detected using ES-EOT (CTRV+ES) as blue surfaces, and the true vehicle skeletons outlined in black (frame 30, 150 and 240, respectively).} 
    \label{fig:change lane track}
  \end{minipage}\hfill
  \begin{minipage}[t]{0.5\textwidth}
    \centering
    \includegraphics[width=\linewidth]{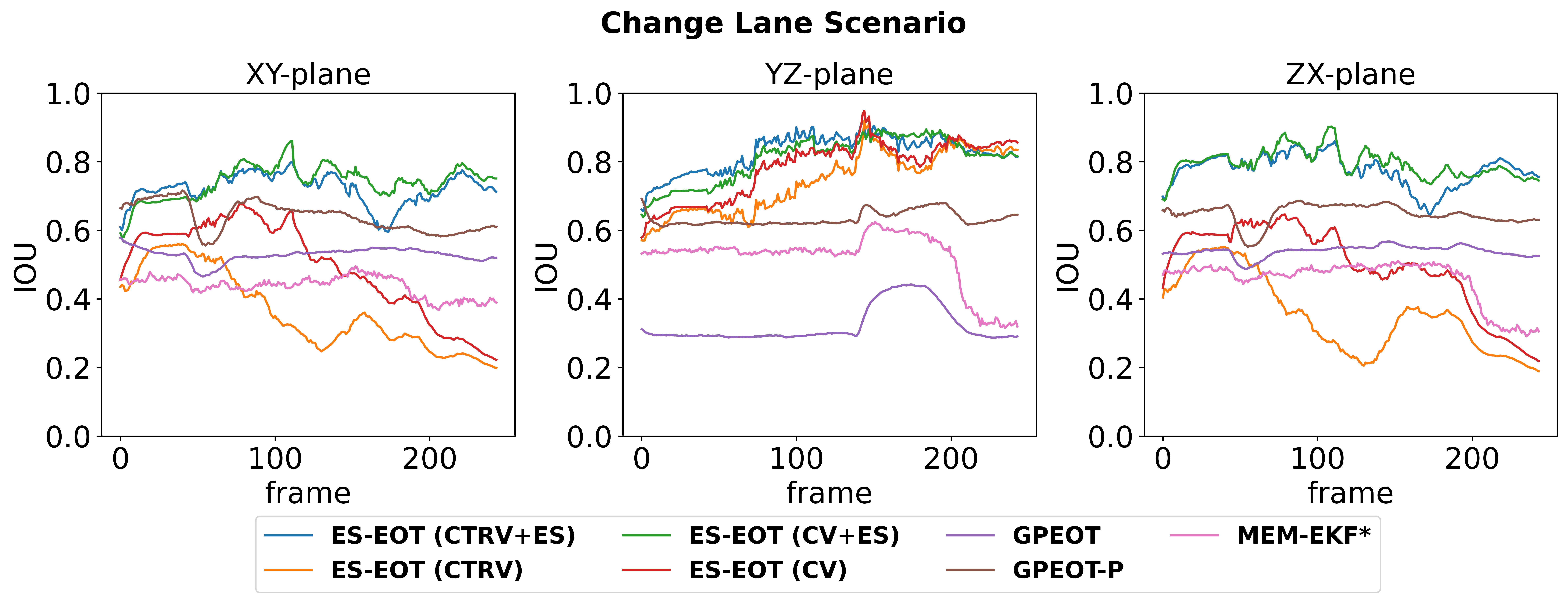}
    \caption{The IOU of the vehicle's shape projections onto the VCS's ``xy-plane'', ``xz-plane'', and ``yz-plane'' in the change lane scenario (averaged over 50 runs with $\alpha = 10$)} 
    \label{fig:change lane iou with frame number}
  \end{minipage}
\end{figure}

\begin{table*}[htbp]
  \footnotesize
  \centering
  \caption{Mean IOU and Velocity RMSE for the Change Lane Scenario}
  \label{table:change lane}
  \begin{tabular}{cccccccccccccc}
    \toprule
    \multirow{2}{*}{Methods} & \multirow{2}{*}{Models} & \multicolumn{3}{c}{$\overline{\text{IOU}}_{\text{xy-plane}} \uparrow$} & \multicolumn{3}{c}{$\overline{\text{IOU}}_{\text{yz-plane}} \uparrow$} & \multicolumn{3}{c}{$\overline{\text{IOU}}_{\text{zx-plane}} \uparrow$} & \multicolumn{3}{c}{$\text{RMSE}_v \downarrow$} \\
    \cmidrule(r){3-14} 
    & & $\alpha = 1$ & $\alpha = 5$ & $\alpha = 10$ & $\alpha = 1$ & $\alpha = 5$ & $\alpha = 10$ & $\alpha = 1$ & $\alpha = 5$ & $\alpha = 10$ & $\alpha = 1$ & $\alpha = 5$ & $\alpha = 10$ \\
    \midrule
    \multirow{4}{*}{ES-EOT} & CTRV+ES & 0.624 & 0.680 & 0.721 & 0.790 & 0.812 & \textbf{0.827} & 0.625 & 0.727 & 0.776 & \textbf{0.809} & 0.811 & 0.813 \\
    & CTRV & 0.252 & 0.307 & 0.364 & 0.731 & 0.736 & 0.743 & 0.248 & 0.301 & 0.354 & \textbf{0.809} & \textbf{0.809} & \textbf{0.811} \\
    & CV+ES & \textbf{0.685} & \textbf{0.736} & \textbf{0.741} & \textbf{0.808} & \textbf{0.819} & 0.810 & \textbf{0.708} & \textbf{0.779} & \textbf{0.795} & 0.851 & 0.856 & 0.861 \\
    & CV & 0.464 & 0.482 & 0.493 & 0.783 & 0.784 & 0.784 & 0.463 & 0.481 & 0.496 & 0.849 & 0.854 & 0.860 \\
    GPEOT &-& 0.380 & 0.501 & 0.528 & 0.245 & 0.303 & 0.323 & 0.392 & 0.516 & 0.539 & 1.952 & 1.149 & 0.991 \\
    GPEOT-P &-& 0.479 & 0.611 & 0.639 & 0.512 & 0.617 & 0.633 & 0.504 & 0.632 & 0.648 & 1.801 & 1.356 & 1.139 \\
    MEM-EKF* &-& 0.412 & 0.437 & 0.437 & 0.498 & 0.525 & 0.520 & 0.432 & 0.457 & 0.455 & 1.781 & 1.460 & 1.337 \\
    \bottomrule
  \end{tabular}
\end{table*}

In this scenario, the bus moves away from the sensor, enters the interaction, and then changes lanes, as depicted in Fig.~\ref{fig:change lane track}.


Fig.~\ref{fig:change lane iou with frame number} shows the corresponding IOU values for the vehicle's shape projects over frame sequence. At around frame 180, during a lane change by the vehicle, the IOU values of our proposed method and MEM-EKF* method decrease, while the GPEOT and GPEOT-P methods remain relatively unaffected. Despite the decrease, our proposed method still outperforms all the other methods in this scenario. Furthermore, the IOU values of ES-EOT (CV+ES) are slightly better than those of ES-EOT (CTRV+ES), which means that in the case of weak maneuvering, the CV motion model can better represent the object's motion characteristics compared to the CTRV motion model. The performance of ES-EOT (CTRV) and ES-EOT (CV) is worse than that of ES-EOT (CTRV+ES) and ES-EOT (CV+ES), respectively, indicating that the ES fusion mechanism has excellent performance in improving object position accuracy.


Table \ref{table:change lane} displays the mean IOU errors and velocity RMSE for the lane change scenario. Our proposed method exhibits significantly higher IOU values and lower velocity RMSE compared to the other methods.

\subsubsection{Scenario of Turning around}

\begin{figure}[tbp]
  \begin{minipage}[t]{0.5\textwidth}
  \centering
  \subfloat[]{
    \includegraphics[width=4.125cm,height=2.5cm]{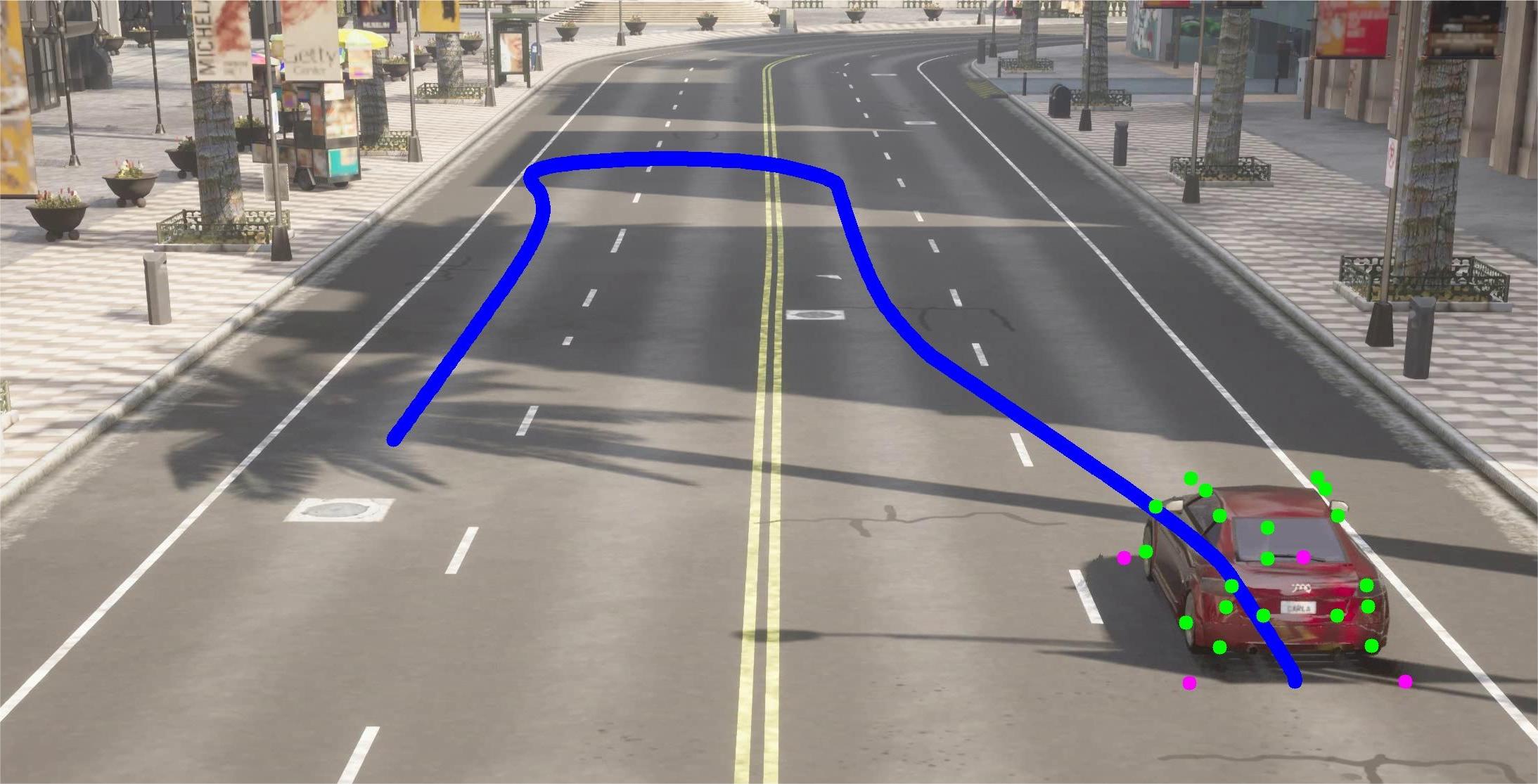}
  }
  \subfloat[]{
    \includegraphics[width=4.125cm,height=2.5cm]{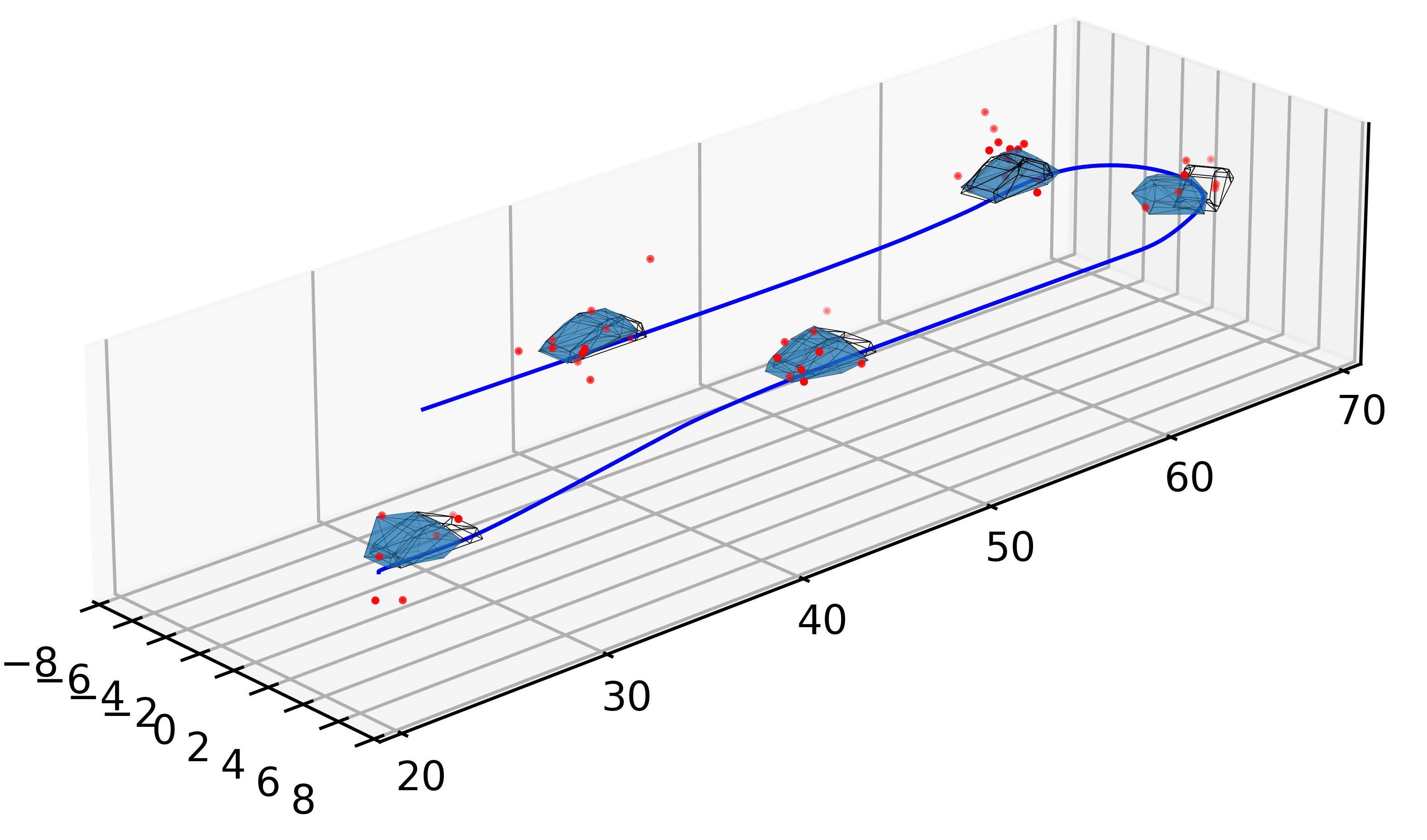}
  }
  \caption{The detections and results from the turning around scenario. The lines and points have the same meaning as depicted in Fig.~\ref{fig:change lane track}. The Fig.~(b) shows frame 30, 90, 150, 180 and 240, respectively.} 
  \label{fig:turn around track}
  \end{minipage}\hfill
  \begin{minipage}[t]{0.5\textwidth}
  \centering
  \includegraphics[width=\linewidth]{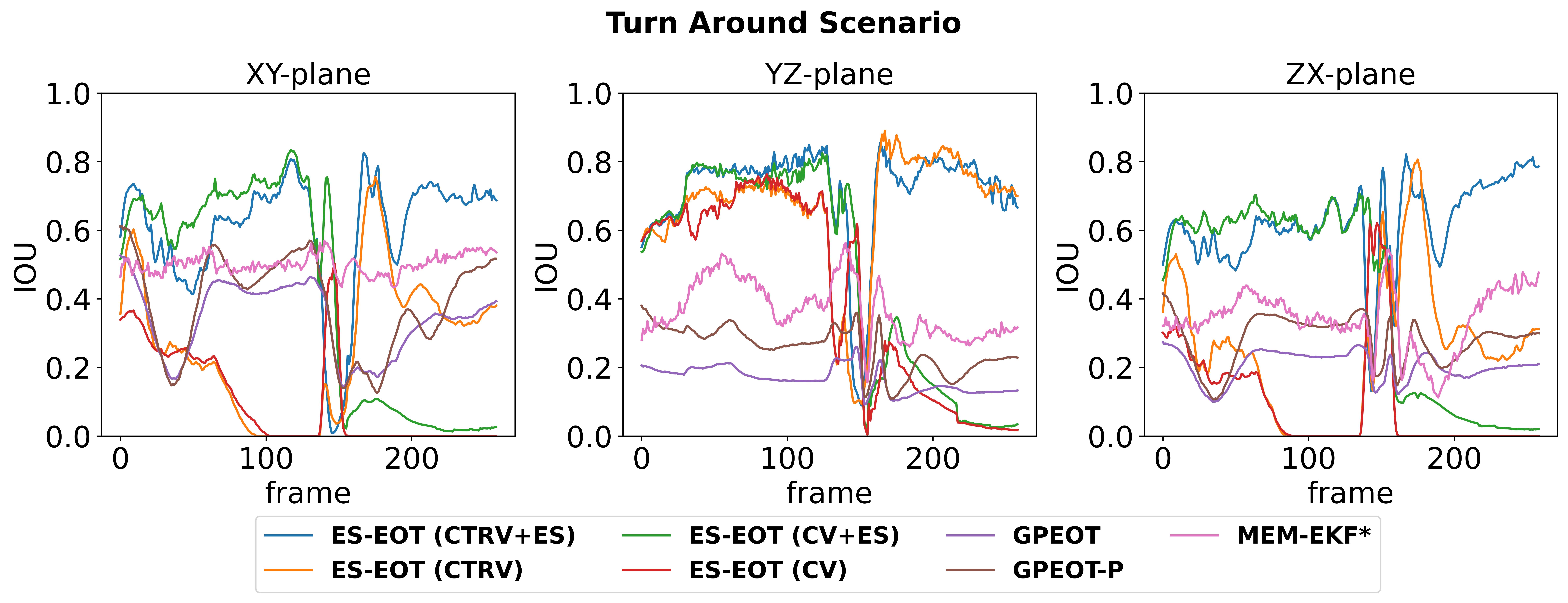}
  \caption{The IOU of the vehicle's shape projections onto the VCS's ``xy-plane'', ``xz-plane'', and ``yz-plane'' in the turn around scenario (averaged over 50 runs with $\alpha = 10$)} 
  \label{fig:turn around iou with frame number}
  \end{minipage}
\end{figure}

In this scenario, the vehicle's motion includes a lane change, a turnaround, and culminating with a precise adjustment to its heading. This intricate sequence of maneuvers is visually depicted in Fig.~\ref{fig:turn around track}.

\begin{table*}[htbp]
  \footnotesize
  \centering
  \caption{Mean IOU and Velocity RMSE for the Turn Around Scenario}
  \label{table:turn around}
  \begin{tabular}{cccccccccccccc}
    \toprule
    \multirow{2}{*}{Methods} & \multirow{2}{*}{Models} & \multicolumn{3}{c}{$\overline{\text{IOU}}_{\text{xy-plane}} \uparrow$} & \multicolumn{3}{c}{$\overline{\text{IOU}}_{\text{yz-plane}} \uparrow$} & \multicolumn{3}{c}{$\overline{\text{IOU}}_{\text{zx-plane}} \uparrow$} & \multicolumn{3}{c}{$\text{RMSE}_v \downarrow$} \\
    \cmidrule(r){3-14} 
    & & $\alpha = 1$ & $\alpha = 5$ & $\alpha = 10$ & $\alpha = 1$ & $\alpha = 5$ & $\alpha = 10$ & $\alpha = 1$ & $\alpha = 5$ & $\alpha = 10$ & $\alpha = 1$ & $\alpha = 5$ & $\alpha = 10$ \\
    \midrule
    \multirow{4}{*}{ES-EOT} & CTRV+ES & \textbf{0.490} & \textbf{0.570} & \textbf{0.602} & \textbf{0.703} & \textbf{0.725} & \textbf{0.716} & \textbf{0.451} & \textbf{0.580} & \textbf{0.622} & \textbf{0.997} & \textbf{0.976} & 0.966 \\
    & CTRV & 0.246 & 0.256 & 0.264 & 0.641 & 0.656 & 0.666 & 0.238 & 0.247 & 0.256 & 1.003 & \textbf{0.976} & \textbf{0.954} \\
    & CV+ES & 0.342 & 0.400 & 0.411 & 0.504 & 0.493 & 0.470 & 0.317 & 0.388 & 0.399 & 4.292 & 4.247 & 4.193 \\
    & CV & 0.087 & 0.096 & 0.101 & 0.427 & 0.418 & 0.410 & 0.100 & 0.100 & 0.100 & 4.290 & 4.246 & 4.194 \\
    GPEOT &-& 0.188 & 0.285 & 0.340 & 0.117 & 0.143 & 0.165 & 0.115 & 0.173 & 0.205 & 2.276 & 1.667 & 1.480 \\
    GPEOT-P &-& 0.282 & 0.350 & 0.386 & 0.230 & 0.238 & 0.255 & 0.234 & 0.260 & 0.280 & 1.560 & 1.410 & 1.344 \\
    MEM-EKF* &-& 0.250 & 0.475 & 0.501 & 0.157 & 0.331 & 0.369 & 0.150 & 0.308 & 0.342 & 1.349 & 1.178 & 1.107 \\
    \bottomrule
  \end{tabular}
\end{table*}

Fig.~\ref{fig:turn around iou with frame number} illustrates the IOU values for the vehicle's shape projections over frame sequence. The IOU values decrease sharply in three places, each corresponding to a vehicle turn. At around frame 50, our proposed method and the MEM-EKF* method are almost unaffected by the lane change, whereas the GPEOT and GPEOT-P methods exhibit significant performance decline. At around frame 150, though all methods' performance degrades significantly during a vehicle turnaround, our method quickly recovers accurate tracking. However, ES-EOT (CV+ES) and ES-EOT (CV) fail after the vehicle turns around, indicating it is unsuitable for complex maneuvering situations. Near frame 200, as the vehicle adjusts its heading direction, the performance of all methods decreases slightly. 


In summary, our proposed method shows strong adaptability to complex movements during the entire tracking process. Furthermore, Table \ref{table:turn around} presents the mean IOU errors and velocity RMSE for the turn around scenario, consistently illustrating our proposed method's superior performance compared to existing methods within complex maneuvering scenarios.

\subsection{Experimental Results}

We also evaluated our proposed method in a challenging scenario using the nuScenes dataset \cite{caesar2020nuscenes}, and compared with the state-of-the-art deep learning based radar camera fusion methods named CRN \cite{kim2023crn} and HVDetFusion \cite{lei2023hvdetfusion}. In this scenario, the ego vehicle remained stationary while the SUV we tracked turned left from a position approximately 15 meters away from the ego vehicle to a position approximately 45 meters away\footnote{Python implementation and data of the methods on nuScenes test scenario are available at https://github.com/RadarCameraFusionTeam-BUPT/ES-EOT-real-nuscenes}. 

The sampling rate in the nuScenes dataset is 0.5Hz, which is relatively low. To adapt to this low sampling rate, we adjusted the prediction uncertainty $\mathbf W$ to $\mathrm{diag}\left( \mathbf 0_{3 \times 3}, 0.1, \mathbf 0_{3 \times 3}, \mathbf I_{3 \times 3}, \mathbf I_{2 \times 2} \right)$. Additionally, we adjusted the hyper-parameters of the ground motion constraint $\mathbf Q^{grnd}$ to $10^{-3}$ to account for the estimation error of the road plane equation, while keeping the other hyper-parameters unchanged from the simulation scenarios.

\begin{figure}[tbp]
  \centering
  \includegraphics[width=0.48\textwidth]{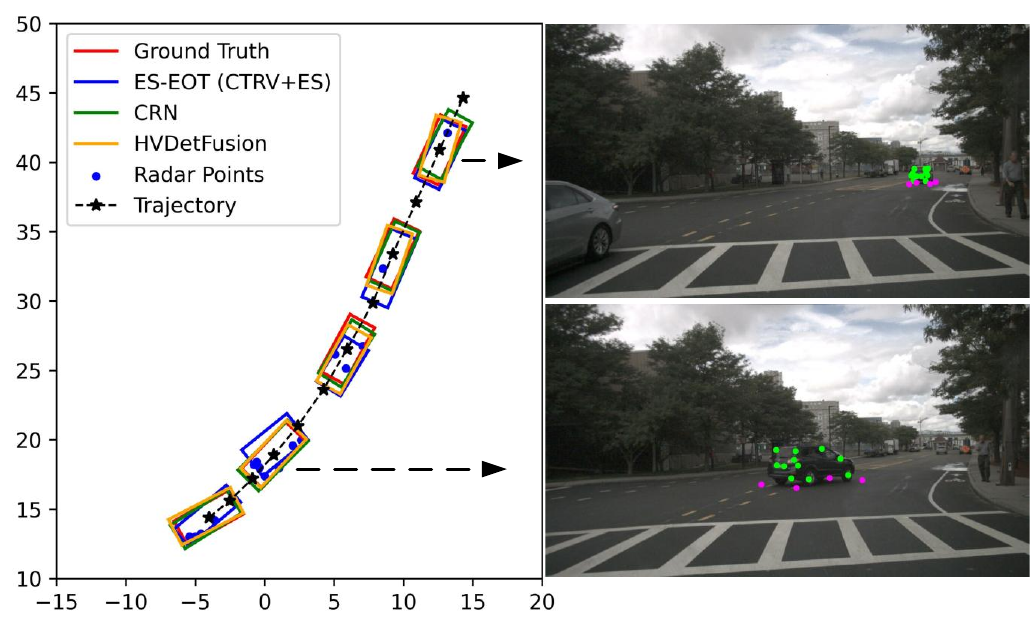}
  \vspace{5pt}
  \caption{Bird-eye-view of the tracking results with nuScenes dataset, and the keypoints detections.}
  \label{fig:BEV}
\end{figure}

Given that the nuScenes dataset provides object labels in the form of 3D bounding boxes rather than skeletons, we utilize the evaluation metrics Average Translation Error (ATE) and Average Scale Error (ASE) provided by the dataset to assess the methods. As shown in Fig.~\ref{fig:BEV}, our proposed method effectively captures the sparse and non-uniform distribution characteristics of radar point clouds, seamlessly integrating with visual measurements to provide accurate estimations of position and size. Notably, our method necessitates no further training on the new dataset. The only adjustments needed are the road plane equation and some parameters described in the above paragraph. 

Table \ref{table:ATE and ASE} shows the ATE and ASE values of the methods. Although our proposed method may not surpass the performance of deep learning based methods in vehicle-mounted scenarios, it still achieves satisfactory accuracy. Importantly, our proposed method is better suited for applications in roadside scenes, especially in situations with limited pixel annotations or even no annotations.

\begin{table}[htbp]
  \centering
  \caption{ATE and ASE For the Experimental Scene}
  \label{table:ATE and ASE}
  \begin{tabular}{c|c|c|c}
    \hline
    & Trained & $\text{ATE} \downarrow$ & $\text{ASE} \downarrow$ \\
    \hline
    ES-EOT (CTRV+ES) & no & 0.760 & 0.175 \\
    \hline
    CRN & yes & \textbf{0.228} & \textbf{0.060} \\
    \hline
    HVDetFusion & yes & 0.337 & 0.063 \\
    \hline
  \end{tabular}
\end{table}

\subsection{Discussion}

The proposed CTRV motion model, SGW-GMM radar measurement model, and ES fusion mechanism are all manually designed models for recursive Bayesian estimate. This means that our method can be directly applied to new scenes by changing a few parameters without requiring additional training or manual annotations. This is beneficial to large-scale roadside deployment without annotations. Furthermore, our method has strong interpretability and offers valuable insights for heterogeneous data fusion.

However, despite these advantages, our method does have certain limitations. For instance, the measurement models assume Gaussian or Gaussian mixture distributions. Consequently, if the actual measurements deviate from these distributions, the performance of our method may suffer a significant decline. Additionally, during complex maneuvers of the object, there might be a temporary decline in algorithm performance due to the high degree of freedom in the extended state. Moreover, although our proposed method can be directly applied to vehicle perception scenarios, the performance in such scenarios may be reduced compared to roadside perception scenarios, mainly due to severe self-occlusion of vehicles.

\section{Conclusion and Future Work}

This paper introduces a novel 3D EOT method tailored for roadside scenarios, effectively combining radar point clouds and vision pixel keypoints. We propose a spherical Gaussian weighted radar measurement model to handle non-uniform radar point clouds. To cope with sparse point clouds, we propose an elastic skeleton model that fuses vision pixel keypoints with 3D radar point clouds, using the positions of point clouds as scale factors. This approach ensures both precise depiction of the vehicle shape and an accurate determination of the vehicle's position, even in the presence of sparse point clouds. Additionally, we extend the existing 2D constant turn rate and velocity model to a 3D setting to accommodate vehicle motion. Through simulations and experiments in diverse scenarios, our method demonstrates effectiveness in handling sparse and non-uniform point clouds, as well as complex vehicle maneuvers.

For future work, we will consider the effect of sensing range on resolution in simulation and evaluate our methods on more public datasets, such as VoD dataset \cite{palffy2022multi}.

\bibliographystyle{IEEEtran}
\bibliography{reference}

\clearpage 
\begin{center}
  \textbf{\LARGE Supplementary Materials}
\end{center}

\section*{Appendix A\\Derivation For the Time Update}

We provide the derivations for the time update of the reference-state and error-state kinematics that are involved in \eqref{eqn:state predict1}-\eqref{eqn:Sigma_0-2}.

\subsection{Time Update for Reference-state Kinematics}

From \eqref{reference-state kinematic}, it is established that velocity, angular velocity, and the vehicle's dimensions remain constant during $\Delta t$. Consequently, \eqref{eqn:v_ref}, \eqref{eqn:omega_ref} and \eqref{eqn:xi_ref} are self-evident, while derivations are needed for \eqref{eqn:p_ref} and \eqref{eqn:q_ref}. In this section, we simplify notation by omitting the discrete time index $k$ and the subscript `ref'. Additionally, we introduce a continuous-time variable $t$ for clearer illustration of states throughout the time period.

\subsubsection{Time Update for Position State} \label{sec: Time Update for Position State}
We first prove that the rotation matrix format of the vehicle's pose at time $t$ is
\begin{equation}
    \label{eqn: R_t}
    \mathbf R\left( t \right) = \exp \left( [\bm \omega t]_\times \right) \mathbf R\left( 0 \right) \text{, }
\end{equation}
where $\mathbf R \left( 0 \right) = \mathbf R \{ \tilde{\mathbf q}_{ref,k} \}$ represents the rotation matrix format of vehicle's pose at last time step, and the definition of $\exp \left( \cdot \right)$ is in \eqref{eqn:R_theta}. We will continue to demonstrate the correctness of \eqref{eqn: R_t} by proving that its time derivative is equivalent to \eqref{eqn: R_ref_derivative}
\begin{align}
    \label{eqn: partial R_t}
    \frac{\partial \mathbf R\left( t \right)}{\partial t} =&\text{ } \frac{\partial \exp \left( [\bm \omega t]_\times \right)}{\partial t} \mathbf R\left( 0 \right) \notag \\
    =& \left( \cos \left( \| \bm \omega \| t \right) \| \bm \omega \| \left[ \frac{\bm \omega}{\| \bm \omega \|} \right]_\times \right. \notag \\
    &+ \left. \sin \left( \| \bm \omega \| t \right) \| \bm \omega \| \left[ \frac{\bm \omega}{\| \bm \omega \|} \right]^2_\times \right) \mathbf R\left( 0 \right) \notag \\
    =& \left( \cos \left( \| \bm \omega \| t \right) \left[ \bm \omega \right]_\times + \frac{\sin \left( \| \bm \omega \| t \right)}{\| \bm \omega \|} \left[ \bm \omega \right]^2_\times \right) \mathbf R\left( 0 \right) \text{, }
\end{align}
where the $\left[ \bm \omega \right]_\times$ is a skew-symmetric matrix that represents the cross-product of the $\bm \omega$, which is defined in \eqref{eqn: skew matrix}. The third power of a skew-symmetric matrix has the following property:
\begin{equation}
    \label{eqn: third power}
    \left[ \bm \omega \right]_\times^3 = - \| \bm \omega \|^2 \left[ \bm \omega \right]_\times \text{. }
\end{equation}
According to this property, \eqref{eqn: partial R_t} can be simplified as
\begin{subequations}
\begin{align}
    \frac{\partial \mathbf R\left( t \right)}{\partial t} =& \text{ } \left[ \bm \omega \right]_\times \exp \left( [\bm \omega t]_\times \right) \mathbf R\left( 0 \right) \text{, } \\
    \dot{\mathbf R}(t)=& \text{ } \left[ \bm \omega \right]_\times \mathbf R\left( t \right) \label{eqn: derivative of R_t}
\end{align}
\end{subequations}
The expression in \eqref{eqn: derivative of R_t} is equivalent to \eqref{eqn: R_ref_derivative}. Thus, \eqref{eqn: R_t} is correct. According to this, the time update for position state can be obtained by integrating \eqref{reference-state p}
\begin{align}
    \label{eqn: continuous time p}
    \mathbf p(t) =& \text{ } \mathbf p(0) + v \times \int_{0}^{t} \mathbf R(t) dt \mathbf u^d \notag \\
    =& \text{ } \mathbf p(0) + v \Bigg\{ \mathbf I \times t + \frac{1}{\left\| \bm \omega \right\|} \left[ \frac{\bm \omega}{\left\| \bm \omega \right\|} \right]_\times \left( 1- \cos \left( \left\| \bm \omega \right\| t \right) \right) \notag \\
    &+ \left[ \frac{\bm \omega}{\left\| \bm \omega \right\|} \right]_\times^2 \left( t - \frac{\sin \left( \left\| \bm \omega \right\| t \right)}{\left\| \bm \omega \right\|} \right) \Bigg\} \cdot \mathbf R(0) \mathbf u^d \text{, }
\end{align}
where $\mathbf p(0)$ denotes the vehicle's position at the previous time step, implying that $\mathbf p(0) = \mathbf p_{\text{ref}, k}$. Consequently, \eqref{eqn:p_ref} is correct when $t = \Delta t$ in \eqref{eqn: continuous time p}.

\subsubsection{Time Update for Pose State} \label{sec: Time Update for Pose State}
We assume that the reference-state of pose at time $t$ is $\tilde{\mathbf q}(t)$, and $\tilde{\mathbf q}(0)$ is equal to the state at the last time step, which means
\begin{equation}
    \tilde{\mathbf q}(0) = \begin{bmatrix} q_w(0) & \mathbf q_v^{\text{T}}(0) \end{bmatrix}^{\text{T}} = \tilde{\mathbf q}_{ref,k} \text{. }
\end{equation}
Thus, the continuous-time format of \eqref{eqn:q_ref} can be written as
\begin{equation}
    \label{eqn: continuous time q}
    \tilde{\mathbf q}(t) = \mathrm{Exp} \left( \bm \omega \times t \right) \odot \tilde{\mathbf q}(0) \text{. }
\end{equation}

Similar to Section \ref{sec: Time Update for Position State}, we will demonstrate the correctness of \eqref{eqn: continuous time q} by proving that its time derivative is equivalent to \eqref{eqn: dot_q}. According to the definitions of $\mathrm{Exp}(\cdot)$ in \eqref{eqn: Exp and Log}, the vector form of $\mathrm{Exp}(\bm \omega \times t)$ is
\begin{equation}
    \mathrm{Exp}(\bm \omega \times t) = \begin{bmatrix} \cos \left( \frac{\| \bm \omega \| t}{2} \right)  \\ \frac{\bm \omega}{\| \bm \omega \|} \sin \left( \frac{\| \bm \omega \| t}{2} \right) \end{bmatrix} \text{. }
\end{equation}
We then expand the right-hand side of \eqref{eqn: continuous time q}
\begin{align}
    \tilde{\mathbf q}(t) =& \begin{bmatrix} q_w(0) \cos \left( \frac{\| \bm \omega \| t}{2} \right) - \frac{\sin \left( \frac{\| \bm \omega \| t}{2} \right)}{\| \bm \omega \|} \bm \omega^{\text{T}} \mathbf q_v(0) \vspace{5pt}\\ 
    \cos \left( \frac{\| \bm \omega \| t}{2} \right) \mathbf q_v(0) + q_w(0) \frac{\sin \left( \frac{\| \bm \omega \| t}{2} \right)}{\| \bm \omega \|} \bm \omega \\
    + \frac{\sin \left( \frac{\| \bm \omega \| t}{2} \right)}{\| \bm \omega \|} \bm \omega \times \mathbf q_v(0) \end{bmatrix} \text{, }
\end{align}
and its time derivative is
\begin{align}
    \frac{\partial \tilde{\mathbf q}(t)}{\partial t} \notag =&\begin{bmatrix} -q_w(0) \sin \left( \frac{\| \bm \omega \| t}{2} \right) \frac{\| \bm \omega \|}{2} - \frac{\cos \left( \frac{\| \bm \omega \| t}{2} \right)}{2} \bm \omega^{\text{T}} \mathbf q_v(0) \\ 
    \\
    -\frac{\| \bm \omega \| \sin \left( \frac{\| \bm \omega \| t}{2} \right)}{2} \mathbf q_v(0) + q_w(0) \frac{\cos \left( \frac{\| \bm \omega \| t}{2} \right)}{2} \bm \omega \\
    + \frac{\cos \left( \frac{\| \bm \omega \| t}{2} \right)}{2} \bm \omega \times \mathbf q_v(0) \end{bmatrix} \notag \\
    =& \frac{1}{2} \begin{bmatrix} 0 \\ \bm \omega \end{bmatrix} \odot \left( \mathrm{Exp}(\bm \omega \times t) \odot \tilde{\mathbf q}(0) \right) \notag \\
    =& \frac{1}{2} \tilde{\bm \omega} \odot \tilde{\mathbf q}(t)
    \text{, }
\end{align}
which can be rewritten as
\begin{equation}
    \label{eqn: time derivative of pose}
    \dot{\tilde{\mathbf q}} = \frac{1}{2} \tilde{\bm \omega} \odot \tilde{\mathbf q} \text{. }
\end{equation}
The expression in \eqref{eqn: time derivative of pose} is equivalent to \eqref{eqn: dot_q}. As a result, \eqref{eqn: continuous time q} is verified, affirming the validity of \eqref{eqn:q_ref} when $t = \Delta t$.

\subsection{Time Update for Error-state Kinematics}

\subsubsection{Proof of \eqref{error-state kinematic}} \label{subsubsection: Proof of error-state kinematic}
To prove the correctness of \eqref{error-state kinematic}, we subtract the reference-state from \eqref{true kinematic model} and discard second-order small quantity. Since the time derivatives of $v_{ref}$, $\bm \omega_{ref}$ and $\bm \xi_{ref}$ are zero, the time derivations of the corresponding error-state kinematics are straightforward to obtain from \eqref{true kinematic model}. Therefore, we focus on deriving the error-state kinematics for position and pose.

To derive the time derivative of error-state kinematics for position, we first expand \eqref{eqn: dot_p} as reference-state and error-state kinematics
\begin{align}
    \label{eqn: time derivative of p first}
  \dot{\mathbf p} = &\text{ } v \mathbf d \notag \\
  \dot{\mathbf p}_{ref}+\dot{\delta \mathbf p} = & \left( v_{ref}+\delta v \right) \delta \mathbf R \cdot \mathbf R_{ref} \mathbf u^d \notag \\
  = &\text{ } v_{ref} \cdot \delta \mathbf R \cdot \mathbf R_{ref} \mathbf u^d + \delta v \cdot \delta \mathbf R \cdot \mathbf R_{ref} \mathbf u^d \text{. }
\end{align}
According to the definition \eqref{eqn:R_theta} of the transformation from the rotation vector $\bm \theta$ to the rotation matrix $\mathbf R$, a small rotation follows the subsequent property
\begin{equation}
    \label{eqn: small change of R}
    \delta \mathbf R = \exp \left( [\delta \bm \theta]_\times \right) \approx \mathbf I + [\delta \bm \theta]_\times \text{. }
\end{equation}
Substituting \eqref{eqn: small change of R} into \eqref{eqn: time derivative of p first}, the expression can be further simplified by ignoring second-order small quantity
\begin{align}
    \label{eqn: time derivative of p second}
  \dot{\mathbf p}_{ref}+\dot{\delta \mathbf p} \approx &\text{ } v_{ref} \cdot \delta \mathbf R \cdot \mathbf R_{ref} \mathbf u^d + \delta v \cdot \mathbf R_{ref} \mathbf u^d \notag \\
  = &\text{ } v_{ref} \cdot \mathbf R_{ref} \mathbf u^d + v_{ref} \cdot [\delta \bm \theta]_\times \cdot \mathbf R_{ref} \mathbf u^d \notag \\
  & + \delta v \cdot \mathbf R_{ref} \mathbf u^d \text{, }
\end{align}
then we subtract the time derivative of reference-state kinematics \eqref{reference-state p}, and the error-state kinematics for position can be obtained
\begin{align}
  \dot{\delta \mathbf p} = &\text{ } v_{ref} \cdot [\delta \bm \theta]_\times \cdot \mathbf R_{ref} \mathbf u^d + \delta v \cdot \mathbf R_{ref} \mathbf u^d \notag \\
  = &\text{ } -v_{ref} \cdot [\mathbf R_{ref} \mathbf u^d]_\times \delta \bm \theta + \delta v \cdot \mathbf R_{ref} \mathbf u^d \text{. }
\end{align}

Similar to the above derivation, we expand \eqref{eqn: dot_q} as reference-state and error-state kinematics and derive the time derivative of error-state kinematics for pose. We first expand the left side of \eqref{eqn: dot_q}
\begin{equation}
    \label{eqn: expand left side of pose first}
  \dot{\tilde{\mathbf q}} = \dot{\delta \tilde{\mathbf{q}}} \odot \tilde{\mathbf{q}}_{ref} + \delta \tilde{\mathbf{q}} \odot \dot{\tilde{\mathbf{q}}}_{ref} \text{. }
\end{equation}
and \eqref{eqn: expand left side of pose first} can be expanded by substituting reference-state kinematics for pose described in \eqref{reference-state kin q} into it
\begin{equation}
    \label{eqn: eqn: expand left side of pose second}
    \dot{\tilde{\mathbf q}} = \dot{\delta \tilde{\mathbf{q}}} \odot \tilde{\mathbf{q}}_{ref} + \frac{1}{2} \delta \tilde{\mathbf{q}} \odot \tilde{\bm \omega}_{ref} \odot \tilde{\mathbf{q}}_{ref} \text{, }
\end{equation}
Then we expand the right-hand side of \eqref{eqn: dot_q}
\begin{equation}
    \label{eqn: expand right side of pose}
  \frac{1}{2} \tilde{\bm \omega} \odot \tilde{\mathbf q} = \frac{1}{2} \tilde{\bm \omega} \odot \delta \tilde{\mathbf{q}} \odot \tilde{\mathbf{q}}_{ref} \text{. }
\end{equation}
According to the original equation, \eqref{eqn: eqn: expand left side of pose second} is equal to \eqref{eqn: expand right side of pose}
\begin{equation}
    \dot{\delta \tilde{\mathbf{q}}} \odot \tilde{\mathbf{q}}_{ref} + \frac{1}{2} \delta \tilde{\mathbf{q}} \odot \tilde{\bm \omega}_{ref} \odot \tilde{\mathbf{q}}_{ref} = \frac{1}{2} \tilde{\bm \omega} \odot \delta \tilde{\mathbf{q}} \odot \tilde{\mathbf{q}}_{ref} \text{, }
\end{equation}
and the right product of $\tilde{\mathbf{q}}_{ref}$ in both sides of the formula can be extracted \cite[Eq. 16]{sola2017quaternion}
\begin{equation}
    \left( \dot{\delta \tilde{\mathbf{q}}} + \frac{1}{2} \delta \tilde{\mathbf{q}} \odot \tilde{\bm \omega}_{ref} \right) \odot \tilde{\mathbf{q}}_{ref} = \left( \frac{1}{2} \tilde{\bm \omega} \odot \delta \tilde{\mathbf{q}} \right) \odot \tilde{\mathbf{q}}_{ref} \text{, }
\end{equation}
therefore, the equations within the parentheses on both sides are equivalent 
\begin{equation}
    \label{eqn: final time derivation of pose}
    \dot{\delta \tilde{\mathbf{q}}} + \frac{1}{2} \delta \tilde{\mathbf{q}} \odot \tilde{\bm \omega}_{ref} = \frac{1}{2} \tilde{\bm \omega} \odot \delta \tilde{\mathbf{q}} \text{, }
\end{equation}
which can be further reorganized as
\begin{equation}
    \label{eqn: final time derivation of pose rerog}
  2 \dot{\delta \tilde{\mathbf{q}}} = \tilde{\bm \omega} \odot \delta \tilde{\mathbf{q}} - \delta \tilde{\mathbf{q}} \odot \tilde{\bm \omega}_{ref} \text{. }
\end{equation}
where $\delta \tilde{\mathbf{q}}$ can be approximated by a small rotation vector $\delta \bm \theta$
\begin{equation}
    \label{eqn: delta q}
    \delta \tilde{\mathbf{q}} = \mathrm{Exp} \left( \delta \bm \theta \right) \approx \begin{bmatrix}
        1 \\ \frac{1}{2} \delta \bm \theta
    \end{bmatrix} \text{. }
\end{equation}
Then, subtracting \eqref{eqn: delta q} into \eqref{eqn: final time derivation of pose rerog}, we get
\begin{equation}
    \begin{bmatrix}
        0 \\ \frac{1}{2} \dot{\delta \bm \theta}
    \end{bmatrix} \approx \begin{bmatrix} 0 & -\delta \bm \omega^\text{T} \\ \delta \bm \omega & [2\bm \omega_{ref}+\delta \bm \omega]_\times \end{bmatrix} \begin{bmatrix} 1 \\ \delta \bm \theta / 2 \end{bmatrix} \text{. }
\end{equation}
Omitting the second-order small quantity, we obtain
\begin{equation}
  \dot{\delta \bm \theta}=[\bm \omega_{ref}]_\times \delta \bm \theta + \delta \bm \omega \text{. }
\end{equation}

In summary, the matrix form of error-state kinematics can be rewritten as \eqref{error-state kinematic}.

\subsubsection{Proof of \eqref{error-state time update}-\eqref{eqn:Sigma_0-2}}
By integrating \eqref{error-state kinematic}, the time update for error-state kinematics can be expressed as
\begin{equation}
    \label{eqn: exp time update delta x}
  \delta \mathbf x_{k+1} = \exp \left( \mathbf F \Delta t \right) \delta \mathbf x_k + \mathbf w_{k} \text{, }
\end{equation}
where $\mathbf w$ is the same as the definition in \eqref{eqn: delta_x} and
\begin{equation}
  \label{Phi series}
  \exp \left( \mathbf F \Delta t \right) = \mathbf I + \sum_{i = 1}^{\infty} \frac{\left( \mathbf F \Delta t \right)^i}{i!} \text{, } 
\end{equation}
where $\mathbf F$ is defined in \eqref{eqn: F_M0_M1}. 

According to mathematical induction, the exponential of $\mathbf F \Delta t$ has the following properties, when $i \geq 2$
\begin{align}
    \label{eqn: Ft_i}
  \left(\mathbf F \Delta t \right)^i & = \begin{bmatrix} 0 & 0 & \mathbf M_1 \left(\mathbf M_2 \right)^{i-1} & \mathbf M_1 \mathbf M_2^{i-2} & 0\\ 0 & 0 & 0 & 0 & 0 \\ 0 & 0 & \left(\mathbf M_2 \right)^i & \left(\mathbf M_2 \right)^{i-1} & 0 \\ 0 & 0 & 0 & 0 & 0 \\ 0 & 0 & 0 & 0 & 0 \end{bmatrix} \Delta t^i \text{, }
\end{align}
for ease of reference, we define a symbol $\Sigma_n$
\begin{equation}
  \Sigma_n \triangleq \Delta t^n \sum_{k=0}^{\infty} \frac{1}{(k+n)!} \mathbf M_2^k \Delta t^k \text{, }
\end{equation}
then the expression of \eqref{Phi series} can be simplified as
\begin{equation}
    \label{eqn:error-state transition matrix_1}
    \exp \left( \mathbf F \Delta t \right) = \begin{bmatrix} \mathbf I & \mathbf M_0 \Delta t & \mathbf M_1 \Sigma_1 & \mathbf M_1 \Sigma_2 & 0 \\ 0 & \mathbf I & 0 & 0 & 0 \\ 0 & 0 & \Sigma_0 & \Sigma_1 & 0 \\ 0 & 0 & 0 & \mathbf I & 0 \\ 0 & 0 & 0 & 0 & \mathbf I \end{bmatrix} \text{, } 
\end{equation}
which is the same as \eqref{eqn:error-state transition matrix}. We next provide the closed-form of $\Sigma_0$, $\Sigma_1$ and $\Sigma_2$ in \eqref{eqn:error-state transition matrix_1}. When $n = 0$, $\Sigma_0$ is
\begin{equation}
    \Sigma_0 = \sum_{k=0}^{\infty} \frac{1}{k!} \mathbf M_2^k \Delta t^k = \exp \left( \mathbf M_2 \Delta t \right) 
    \text{, }
\end{equation}
which is the same as \eqref{eqn: Sigma_0}. When $n = 1$, $\Sigma_1$ is
\begin{equation}
    \label{eqn: Sigma_1_1}
    \Sigma_1 = \Delta t \sum_{k=0}^{\infty} \frac{1}{(k+1)!} \mathbf M_2^k \Delta t^k 
    \text{, }
\end{equation}
due to that $\mathbf M_2$ is a skew-symmetric matrix, and the third power property described in \eqref{eqn: third power}
\begin{equation}
    \label{eqn: M_2 proper}
    \mathbf M_2 = - \frac{1}{\| \bm \omega_{ref} \|^2} \mathbf M_2^3 \text{. }
\end{equation}
Then substituting \eqref{eqn: M_2 proper} into \eqref{eqn: Sigma_1_1}
\begin{equation}
    \Sigma_1 = \mathbf I \Delta t - \frac{\mathbf M_2}{\|\bm \omega_{ref}\|^2} \left( \Sigma_0 - \mathbf I - \mathbf M_2 \Delta t \right) \text{, }
\end{equation}
which is the same as \eqref{eqn: Sigma_1}, and the similar derivation can be implemented to obtain $\Sigma_2$.

\section*{Appendix B\\Derivation For ES Fusion Mechanism}

In this section, we derive the closed-form of transition model of ES fusion mechanism \eqref{es model time update}-\eqref{eqn: M_3 M_4}. By integrating \eqref{ES dynamic model}, the ES fusing mechanism has the similar form of \eqref{eqn: exp time update delta x}
\begin{equation}
    \bm \vartheta_{k+1, t} = \exp \left( \mathbf F_{\vartheta} \Delta t \right) \bm \vartheta_{k, t} + \mathbf w_{k}^{\vartheta} \text{, }
\end{equation}
where $\mathbf F_{\vartheta}$ and $\mathbf w_{k}^{\vartheta}$ is the same as \eqref{eqn: F_vartheta} and \eqref{eqn: dynamic vartheta}, respectively, and 
\begin{equation}
    \label{eqn: exp vartheta}
  \exp \left( \mathbf F_{\vartheta} \Delta t \right) = \mathbf I + \sum_{i = 1}^{\infty} \frac{\left( \mathbf F_{\vartheta} \Delta t \right)^i}{i!} \text{, }
\end{equation}
following the mathematical induction, we find that
\begin{equation}
  \left( \mathbf F_{\vartheta} \right)^i = \begin{bmatrix} \mathbf 0 & \mathbf 0 & \mathbf 0 \\ -\mathbf M_{3,i} & \mathbf M_{3,i} & \mathbf M_{4,i} \\ \epsilon \mathbf M_{4,i} & -\epsilon \mathbf M_{4,i} & \mathbf M_{3,i} - \rho \mathbf M_{4,i} \end{bmatrix} \text{ , } i \geq 1 \text{, }
\end{equation}
with
\begin{subequations}
  \begin{align}
    \mathbf M_{3,i} & = -\epsilon \mathbf M_{4,i-1} \text{, } \\
    \mathbf M_{4,i} & = -\epsilon \mathbf M_{4,i-2} - \rho \mathbf M_{4,i-1} \text{, }
  \end{align}
\end{subequations}
and
\begin{subequations}
  \begin{align}
    \mathbf M_{4, i} & = -\frac{2^{-i}}{\sqrt{-4\epsilon + \rho^2}} \left( \left( -\rho - \sqrt{-4\epsilon + \rho^2} \right)^i \right. \notag \\
    & - \left. \left( -\rho + \sqrt{-4\epsilon + \rho^2} \right)^i \right) \mathbf I \text{ , } i \geq 1 \text{, }\\
    \mathbf M_{3, i} & = -\epsilon \mathbf M_{4, i-1} \text{ , } i \geq 2 \text{, }
  \end{align}
\end{subequations}
then the closed-form of transition matrix $\Phi_{\vartheta}$ can be obtained by simplifying \eqref{eqn: exp vartheta}, and the results are concluded in \eqref{eqn: M_3 M_4}.

\section*{Appendix C\\Proofs of Lemma 1 to 3}

\setcounter{subsection}{0}

\subsection{Proof of Lemma 1}
The product of a skew-matrix with a vector is equivalent to the cross product, \emph{i.e.},
\begin{equation}
    [\mathbf a]_\times \mathbf b = \mathbf a \times \mathbf b \text{, }
\end{equation}
where $\mathbf a$ and $\mathbf b$ are two arbitury 3D vectors. As a result
\begin{align}
    \label{eqn: lamma 1}
  \left[ \mathbf v \right]_\times \cdot \mathbf M & = \begin{bmatrix} \mathbf v \times \mathbf M_{(1)} & ,\dots, & \mathbf v \times \mathbf M_{(n)} \end{bmatrix} \notag \\
  & = - \begin{bmatrix} \mathbf M_{(1)} \times \mathbf v & ,\dots, & \mathbf M_{(n)} \times \mathbf v \end{bmatrix} \notag \\
  & = -\left[ \mathbf M \right]_* diag_n \left( \mathbf v \right) \text{, }
\end{align}
where $\mathbf{v}$ and $\mathbf{M}$ are a 3D vector  and a $3 \times n$ matrix, respectively. The definition of symbol $\left[ \cdot \right]_\ast$, $diag_n(\cdot)$ and $\mathbf M_{(i)}$ are the same as Lemma 1.

\subsection{Proof of Lemma 2}
By using \eqref{eqn: lamma 1}, Lemma 2 can be written as
\begin{align}
  \left[ \mathbf v \right]_\times \cdot \mathbf M \cdot \mathbf t & = -\left[ \mathbf M \right]_* diag_3 \left( \mathbf v \right) \cdot \mathbf t \notag \\
  & = -\left[ \mathbf M \right]_* \left[ \mathbf v \cdot \mathbf t^\text{T} \right]_{\star} \text{, }
\end{align}
where the definition of $\left[ \cdot \right]_\star$ is the same as Lemma 2.

\subsection{Proof of Lemma 3}
By using \eqref{eqn: lamma 1} and cholesky decomposition, Lemma 3 can be written as
\begin{align}
  & \left[\mathbf v \right]_\times \cdot \mathbf M \cdot \left[\mathbf t \right]_\times^\text{T} \notag \\
  = & \left( \left[ \mathbf v \right]_\times \mathbf L \right) \left( \left[ \mathbf t \right]_\times \mathbf L \right)^\text{T} \notag \\
  = & \left[ \mathbf L \right]_\ast diag_n \left( \mathbf v \right) diag_n \left( \mathbf t^\text{T} \right) \left[ \mathbf L \right]_\ast^\text{T} \notag \\
  = & \left[ \mathbf L \right]_\ast diag_n \left( \mathbf v \cdot \mathbf t^\text{T} \right) \left[ \mathbf L \right]_\ast^\text{T} \text{. }
\end{align}

\section*{Appendix D\\Derivation For $\mathbb{E}_{A}\left[\log p(\Theta_k, A_k, \mathbf Z_k| \mathbf Z^{k-1})\right]$}

We first expand the logarithm of the joint probability density according to \eqref{eqn:likelihood function} and \eqref{eqn: post prob}
\begin{align}
    \label{eqn: log joint prob}
    &\text{ }\log p(\Theta_k, A_k, \mathbf Z_k| \mathbf Z^{k-1}) \notag \\
    =&\text{ } \log p(\Theta_k|\mathbf Z^{k-1}) + \log p(A_k| \Theta_k) + \log p(\mathbf Z_k| \Theta_k, A_k) \notag \\
    =&\text{ } \log p(\Theta_k|\mathbf Z^{k-1}) + \log p(A_k| \Theta_k) + \log p(\mathbf Z_{k}^r|\Theta_{k},A_{k}) \notag \\
    &+ \log p(\mathbf Z_k^c|\Theta_{k}) + \log p\left( z_k^{rot} | \Theta_k \right) + \log p\left( \mathbf Z_k^{grnd} | \Theta_k \right) \notag \\
    &+ \log p\left( \mathbf Z_k^{sym} | \Theta_k \right) \text{, }
\end{align}
where $\log p(A_k| \Theta_k)$ and $\log p(\mathbf Z_{k}^r|\Theta_{k},A_{k})$ are related to $A_k$, while other components are independent of $A_k$. As a result, the expectation of \eqref{eqn: log joint prob} only influences the aforementioned two components, leaving the other parts unchanged
\begin{align}
    \label{eqn: expectation log joint prob A}
    &\text{ }\mathbb{E}_{A}\left[\log p(\Theta_k, A_k, \mathbf Z_k| \mathbf Z^{k-1})\right] \notag \\
    =&\text{ } \mathbb{E}_{A} \left[ \log p(A_k| \Theta_k) + \log p(\mathbf Z_{k}^r|\Theta_{k},A_{k}) \right] \notag \\
    &+\log p(\Theta_k|\mathbf Z^{k-1}) + \log p(\mathbf Z_k^c|\Theta_{k}) + \log p\left( z_k^{rot} | \Theta_k \right) \notag \\
    &+ \log p\left( \mathbf Z_k^{grnd} | \Theta_k \right) + \log p\left( \mathbf Z_k^{sym} | \Theta_k \right) \text{. }
\end{align}
Therefore, we focus on the simplification of the affected components, and expand it according to \eqref{eqn: radar likelihood} and \eqref{eqn: association likelihood}.
\begin{align}
    \label{eqn: E_a affect part}
    &\text{ } \mathbb{E}_{A} \left[ \log p(A_k| \Theta_k) + \log p(\mathbf Z_{k}^r|\Theta_{k},A_{k}) \right] \notag \\
    =&\text{ } \sum_{t=1}^{T} \sum_{u=1}^{n_k^r}\mathbb{E}_{A} \left[ a_{it} \right] \bigg( \log \left( \pi_{k,t}^{(\iota)} \right) \notag \\
    &- \frac{1}{2} \left( \mathbf z_{k,i}^r - \bm \zeta_{k,t} \right)^{\text{T}} \mathbf Q^{-1} \left( \mathbf z_{k,i}^r - \bm \zeta_{k,t} \right) \bigg) + c_{A^*} \text{, }
\end{align}
where $\mathbb{E}_{A} \left[ a_{it} \right] = \upsilon_{it}^{(\iota)}$ at the $\iota$-th iteration, according to \eqref{eqn: update association}. For simplicity, we assume that $\pi_{k,t}^{(\iota)}$ is not a random variable, and its logarithm is calculated based on the reference-states in every iteration, which means that it is a constant towards $\mathbf x_k$ and $\bm \vartheta_k$. Consequently, \eqref{eqn: E_a affect part} can be reformulated as
\begin{align}
    \label{eqn: expectation sim for a}
    &\text{ } \mathbb{E}_{A} \left[ \log p(A_k| \Theta_k) + \log p(\mathbf Z_{k}^r|\Theta_{k},A_{k}) \right] \notag \\
    =&\text{ } -\frac{1}{2} \sum_{t=1}^{T} \sum_{u=1}^{n_k^r} \upsilon_{it}^{(\iota)} \left( \mathbf z_{k,i}^r - \bm \zeta_{k,t} \right)^{\text{T}} \mathbf Q^{-1} \left( \mathbf z_{k,i}^r - \bm \zeta_{k,t} \right) + c \notag \\
    =& \text{ } -\frac{1}{2} \sum_{t=1}^{T} \sum_{u=1}^{n_k^r} \mathrm{tr} \left\{ \upsilon_{it}^{(\iota)} \left( \mathbf z_{k,i}^r - \bm \zeta_{k,t} \right) \left( \mathbf z_{k,i}^r - \bm \zeta_{k,t} \right)^{\text{T}} \mathbf Q^{-1} \right\} + c \notag \\
    =& \text{ } -\frac{1}{2} \mathrm{tr} \left\{ \sum_{t=1}^{T} \sum_{u=1}^{n_k^r} \upsilon_{it}^{(\iota)} \left( \mathbf z_{k,i}^r - \overline{\mathbf z}^{(\iota+1)}_{k,t} + \overline{\mathbf z}^{(\iota+1)}_{k,t} - \bm \zeta_{k,t} \right) \right.\notag \\
    &\cdot \left. \left( \mathbf z_{k,i}^r - \overline{\mathbf z}^{(\iota+1)}_{k,t} + \overline{\mathbf z}^{(\iota+1)}_{k,t} - \bm \zeta_{k,t} \right)^{\text{T}} \mathbf Q^{-1} \right\} + c \notag \\
    =& \text{ } -\frac{1}{2} \mathrm{tr} \Bigg\{ \sum_{t=1}^{T} \sum_{u=1}^{n_k^r} \upsilon_{it}^{(\iota)} \left( \mathbf z_{k,i}^r - \overline{\mathbf z}^{(\iota+1)}_{k,t} \right) \left( \mathbf z_{k,i}^r - \overline{\mathbf z}^{(\iota+1)}_{k,t} \right)^{\text{T}} \mathbf Q^{-1} \notag \\
    &+ \upsilon_{it}^{(\iota)} \left( \overline{\mathbf z}^{(\iota+1)}_{k,t} - \bm \zeta_{k,t} \right) \left( \overline{\mathbf z}^{(\iota+1)}_{k,t} - \bm \zeta_{k,t} \right)^{\text{T}} \mathbf Q^{-1} \Bigg\} + c \notag \\
    =& \text{ } -\frac{1}{2} \sum_{t=1}^{T} \Bigg\{ \mathrm{tr} \Big( \overline{\mathbf Z}^{(\iota+1)}_{k,t} \mathbf Q^{-1} \Big) \notag \\ 
    &+ \left( \overline{\mathbf z}^{(\iota+1)}_{k,t} - \bm \zeta_{k,t} \right)^{\text{T}} \left( \frac{\mathbf Q}{n^{(\iota+1)}_{k,t}} \right)^{-1} \left( \overline{\mathbf z}^{(\iota+1)}_{k,t} - \bm \zeta_{k,t} \right) \Bigg\}  + c
    \text{, } 
\end{align}
where $c$ represents a constant independent to $A_k$, $\mathbf x_k$ and $\bm \vartheta_k$, and the definitions of $n^{(\iota+1)}_{k,t}$, $\overline{\mathbf z}^{(\iota+1)}_{k,t}$ and $\overline{\mathbf Z}^{(\iota+1)}_{k,t}$ are the same as \eqref{eqn:n_kt_Z_line}. In summary, $\mathbb{E}_{A}\left[\log p(\Theta_k, A_k, \mathbf Z_k| \mathbf Z^{k-1})\right]$ can be obtained by substituting \eqref{eqn: expectation sim for a} into \eqref{eqn: expectation log joint prob A}.

\section*{Appendix E\\Closed-Form Expressions for the Posterior Probability Densities of $\mathbf x$ and $\bm \vartheta$}

The posterior probability densities of $\mathbf x$ and $\bm \vartheta$ are in \eqref{eqn: update_state}, where each of the parameters of the densities function consists of six parts, with five corresponding to the likelihood functions and one corresponding to the prior. The parameters in \eqref{eqn: update_x} are as follows
\begin{align}
    \hat{\mathbf P}_{k|k}^{(\iota+1)} =& \Bigg\{ \sum_{t=1}^T  \left( \left(\mathbf H_{1,t,\mathbf x_{k}}^{(\iota)}\right)^T \left( \frac{\mathbf Q}{n_{k,t}^{(\iota+1)}} \right)^{-1} \mathbf H_{1,t,\mathbf x_{k}}^{(\iota)} \right. \notag \\
    &+ \left. \left( \mathbf H^{\theta} \right)^T \left(\mathbf Q_{t,\mathbf x_{k}}^{(\iota)}\right)^{-1} \mathbf H^{\theta} \right) + \left(\hat{\mathbf P}_{k|k}^{(\iota)} \right)^{-1} \notag \\
    &+ \sum_{i=1}^{|\mathbf Z_k^{c, b}|} \left( \mathbf H_{k,\mathbf x}^{c,b,(\iota)} \right)^T \left( \mathbf Q_{k,s_i^b}^{c, b} \right)^{-1} \mathbf H_{k,\mathbf x}^{c,b,(\iota)} \notag \\
    &+ \sum_{i=1}^{|\mathbf Z_k^{c, g}|} \left( \mathbf H_{k,\mathbf x}^{c,g,(\iota)} \right)^T \left( \mathbf Q_{k,s_i^g}^{c, g} \right)^{-1} \mathbf H_{k,\mathbf x}^{c,g,(\iota)} \notag \\
    &+ \left( \mathbf H_{5, \mathbf x_{k}}^{(\iota)} \right)^T \left( \mathbf Q^{rot} \right)^{-1} \mathbf H_{5, \mathbf x_{k}}^{(\iota)} \notag \\
    &+ \sum_{i=1}^{4} \left( \mathbf H_{6, i, \mathbf x_{k}}^{(\iota)} \right)^T \left( \mathbf Q^{grnd} \right)^{-1} \mathbf H_{6, i, \mathbf x_{k}}^{(\iota)} \Bigg\}^{-1} \text{, } \\
    \delta \hat{\mathbf x}_k^{(\iota+1)} = & \hat{\mathbf P}_{k|k}^{(\iota+1)} \Bigg\{ \sum_{t=1}^T  \left ( \left(\mathbf H_{1,t,\mathbf x_{k}}^{(\iota)}\right)^T \left( \frac{\mathbf Q}{n_{k,t}^{(\iota+1)}} \right)^{-1} \mathbf h_{1,t,\mathbf x_{k}}^{(\iota)} \right. \notag \\
    &+ \left. \left( \mathbf H^{\theta} \right)^T \left(\mathbf Q_{t,\mathbf x_{k}}^{(\iota)}\right)^{-1} \mathbf h_{2,t,\mathbf x_{k}}^{(\iota)} \right) \notag \\
    &+ \sum_{i=1}^{|\mathbf Z_k^{c, b}|} \left( \mathbf H_{k,\mathbf x}^{c,b,(\iota)} \right)^T \left( \mathbf Q_{k,s_i^b}^{c, b} \right)^{-1} \mathbf h_{3, s_i^b, \mathbf x_{k}}^{(\iota)} \notag \\
    &+ \sum_{i=1}^{|\mathbf Z_k^{c, g}|} \left( \mathbf H_{k,\mathbf x}^{c,g,(\iota)} \right)^T \left( \mathbf Q_{k,s_i^g}^{c, g} \right)^{-1} \mathbf h_{4, s_i^g, \mathbf x_{k}}^{(\iota)}  \notag \\
    &+ \left( \mathbf H_{5, \mathbf x_{k}}^{(\iota)} \right)^T \left( \mathbf Q^{rot} \right)^{-1} \mathbf h_{5,\mathbf x_{k}}^{(\iota)} \notag \\
    &+ \sum_{i=1}^{4} \left( \mathbf H_{6, i, \mathbf x_{k}}^{(\iota)} \right)^T \left( \mathbf Q^{grnd} \right)^{-1} \mathbf h_{6, i, \mathbf x_{k}}^{(\iota)} \Bigg\} \text{, }
  \end{align}
  and
  \begin{equation}
    \hat{\mathbf x}_{k|k}^{(\iota+1)} = \mathbf x_{ref,k}^{(\iota)} + \delta \hat{\mathbf x}_k^{(\iota+1)} \text{, }
  \end{equation}
  with
  \begin{subequations}
  \begin{align}
    \widetilde{\mathbf J}_{k}^{(\iota)} =& \left[\sqrt{ \left( \frac{\mathbf Q}{n_{k,t}^{(\iota+1)}} \right)^{-1} } \right]_*^T \mathbf J_{k}^{(\iota)} \text{, } \\
    \mathbf H_{1,t,\mathbf x_{k}}^{(\iota)} =& \mathbf H^r - \left[\mathbf R^{(\iota)}_k \mathbb{E} \left[\mathbf u^{(\iota)}_{k,t} \right] \right]_\times \mathbf J_{k}^{(\iota)} \mathbf H^{\theta} \text{, } \\
    \mathbf h_{1,t,\mathbf x_{k}}^{(\iota)} =& \overline{\mathbf z}_{k,t}^{(\iota)} - \mathbf R^{(\iota)}_k \mathbb{E} \left[\mathbf u^{(\iota)}_{k,t} \right] - \mathbf p_{ref, k}^{(\iota)} \text{, } \\
    \left(\mathbf Q_{t,\mathbf x_{k}}^{(\iota)}\right)^{-1} =& \widetilde{\mathbf J}_{k}^{(\iota),T} \mathrm{diag}_3\left( \mathbf R^{(\iota)}_k \mathrm{Cov} \left[ \mathbf u^{(\iota)}_{k,t} \right] \mathbf R^{(\iota),T}_k \right) (\cdot) \text{, } \\
    \mathbf h_{2,t,\mathbf x_{k}}^{(\iota)} =& \mathbf Q_{t,\mathbf x_{k}}^{(\iota),T} \mathbf J_{k}^{(\iota),T} \cdot \left[ \left( \frac{\mathbf Q}{n_{k,t}^{(\iota+1)}} \right)^{-1} \right]_* \notag \\
    &\cdot \left[ \mathbf R^{(\iota)}_k \mathrm{Cov} \left[ \mathbf u^{(\iota)}_{k,t} \right] \mathbf R^{(\iota),T}_k \right]_\star \text{, } \\
    \mathbf h_{3,s_i^b,\mathbf x_{k}}^{(\iota)} =& \mathbf z_{k,i}^{c, b} - h^{c, b}\left( \mathbf x_{ref,k}^{(\iota)}, \bm \varphi_{k,s_i^b}^{(\iota)} \right) \text{, } \\
    \mathbf h_{4,s_i^g,\mathbf x_{k}}^{(\iota)} =& \mathbf z_{k,i}^{c, g} - h_{s_i^g}^{c, g}\left( \mathbf x_{ref,k}^{(\iota)} \right) \text{, } \\
    \mathbf H_{5, \mathbf x_{k}}^{(\iota)}=& \left( \mathbf R^{(\iota)}_k \mathbf u^d \right)^T \mathbf H^{\omega} \notag \\
    &- \bm \omega_{ref, k}^{(\iota), T} \left[ \mathbf R^{(\iota)}_k \mathbf u^d \right]_\times \mathbf J_{k}^{(\iota)} \mathbf H^{\theta} \text{, } \\
    \mathbf h_{5,\mathbf x_{k}}^{(\iota)} =& z_k^{rot} - \left( \mathbf R^{(\iota)}_k \mathbf u^d \right)^T \bm \omega_{ref, k}^{(\iota)} \text{, } \\
    \mathbf H_{6, i, \mathbf x_{k}}^{(\iota)} =& \mathbf n^T \left( \mathbf H^{r} - \left[ \mathbf R_k^{(\iota)} \mathbf G_i \bm \xi_{ref}^{(\iota)} \right]_\times \mathbf J_k^{(\iota)} \mathbf H^{\theta} \right. \notag \\
    &+ \left. \mathbf R_k^{(\iota)} \mathbf G_i \mathbf H^{\xi} \right) \text{, } \\
    \mathbf h_{6, i, \mathbf x_{k}}^{(\iota)} =& - \mathbf n^T \left( \mathbf p_{ref}^{(\iota)} + \mathbf R_k^{(\iota)} \mathbf G_i \bm \xi_{ref}^{(\iota)} \right) - d \text{, }\\
    \mathbf H^{r} =& \begin{bmatrix} \mathbf I_{3 \times 3} & 0 & \mathbf 0 & \mathbf 0 & \mathbf 0 \end{bmatrix} \text{, } \\
    \mathbf H^{\theta} =& \begin{bmatrix} \mathbf 0 & 0 & \mathbf I_{3 \times 3} & \mathbf 0 & \mathbf 0 \end{bmatrix} \text{, } \\
    \mathbf H^{\omega} =& \begin{bmatrix} \mathbf 0 & 0 & \mathbf 0 & \mathbf I_{3 \times 3} & \mathbf 0 \end{bmatrix} \text{, } \\
    \mathbf H^{\xi} =& \begin{bmatrix} \mathbf 0 & 0 & \mathbf 0 & \mathbf 0 & \mathbf I_{2 \times 2} \end{bmatrix}  \text{ }
  \end{align}
\end{subequations}

The parameters in \eqref{eqn: update_vartheta} are as follows
\begin{align}
    \hat{\bm \mu}_{k|k,t}^{(\iota+1)} =& \hat{\bm \Sigma}_{k|k,t}^{(\iota+1)} \Bigg\{ \left( \mathbf H_{1,\bm \vartheta_{k,t}}^{(\iota)} \right)^T \left( \frac{\mathbf Q}{n_{k,t}^{(\iota+1)}} \right)^{-1} \mathbf h_{1,\bm \vartheta_{k,t}}^{(\iota)} \notag \\
    +& \left( \mathbf H^{u} \right)^T \left( \mathbf Q_{\bm \vartheta_{k,t}}^{(\iota)}\right)^{-1} \mathbf h_{2,\bm \vartheta_{k,t}}^{(\iota)} + \left( \hat{\bm \Sigma}_{k|k,t}^{(\iota)} \right)^{-1} \hat{\bm \mu}_{k|k,t}^{(\iota)} \notag \\
    +& \left( \mathbf H_{\varpi} \right)^T \left( \mathbf Q^{sym} \right)^{-1} \mathbf h_{3, \bm \vartheta_{k,t}}^{(\iota)} \notag \\
    +& \sum_{i=1}^{|\mathbf Z_k^{c, b}|} \left( \mathbf H_{4,s_i^b,\bm \vartheta_{k,t}}^{(\iota)} \right)^T \left( \mathbf Q_{k,s_i^b}^{(\iota)} \right)^{-1} \mathbf h_{4,s_i^b,\bm \vartheta_{k,t}}^{(\iota)} \cdot \mathbf 1_t \left( s_i^b \right) \text{, }
  \end{align}
  \begin{align}
    \hat{\bm \Sigma}_{k|k,t}^{(\iota+1)} =& \Bigg\{ \left( \mathbf H_{1,\bm \vartheta_{k,t}}^{(\iota)} \right)^T \left( \frac{\mathbf Q}{n_{k,t}^{(\iota+1)}} \right)^{-1} \mathbf H_{1,\bm \vartheta_{k,t}}^{(\iota)} + \left( \hat{\bm \Sigma}_{k|k,t}^{(\iota)} \right)^{-1} \notag \\
    +& \left( \mathbf H^{u} \right)^T \left( \mathbf Q_{\bm \vartheta_{k,t}}^{(\iota)}\right)^{-1} \mathbf H^u \notag + \left( \mathbf H^{\varpi} \right)^T \left( \mathbf Q^{sym} \right)^{-1} \mathbf H_{\varpi} \notag \\
    +& \sum_{i=1}^{|\mathbf Z_k^{c, b}|} \left( \mathbf H_{4,s_i^b,\bm \vartheta_{k,t}}^{(\iota)} \right)^T \left( \mathbf Q_{k,s_i^b}^{(\iota)} \right)^{-1} \mathbf H_{4,s_i^b,\bm \vartheta_{k,t}}^{(\iota)} \cdot \mathbf 1_t \left( s_i^b \right) \text{, }
  \end{align}
  with
  \begin{subequations}
  \begin{align}
    \widetilde{\mathbf R}^{(\iota),T}_{k,t} =& \left[ \sqrt{\left( \frac{\mathbf Q}{n_{k,t}^{(\iota+1)}} \right)^{-1}} \right]_*^T \mathbf R_k^{(\iota)} \text{, }  \\
    \mathbf H_{1,\bm \vartheta_{k,t}}^{(\iota)} =& \mathbf R_k^{(\iota)} \mathbf H^u \text{, } \\
    \mathbf h_{1,\bm \vartheta_{k,t}}^{(\iota)} =& \overline{\mathbf z}_{k,t}^{(\iota)} - \mathbb{E} \left[ \mathbf p_k^{(\iota)} \right] \text{, } \\
    \left(\mathbf Q_{\bm \vartheta_{k,t}}^{(\iota)}\right)^{-1} =& \widetilde{\mathbf R}^{(\iota),T}_{k,t} diag_3 \left( \mathbf J_{k}^{(\iota)} \mathbb{E} \left[ \delta \bm \theta \delta \bm \theta^T \right]  \mathbf J_{k}^{(\iota),T} \right) (\cdot) \text{, } \\
    \mathbf h_{2,\bm \vartheta_{k,t}}^{(\iota)} =& - \mathbf Q_{\bm \vartheta_{k,t}}^{(\iota),T} \mathbf R_k^{(\iota),T} \cdot \left[ \left( \frac{\mathbf Q}{n_{k,t}^{(\iota+1)}} \right)^{-1} \right]_* \notag \\
    \cdot& \left[ \mathbf J_{k}^{(\iota)} \mathbb{E} \left[ \delta \bm \theta_k^{(\iota)} \delta \mathbf p_k^{(\iota),T} \right] \right]_\star \text{, } \\
    \mathbf h_{3, \bm \vartheta_{k,t}}^{(\iota)} =& \mathbf H^{\varpi} \mathbb{E} \left[ \bm \vartheta_{k,t}^{sym} \right] \text{, } \\
    \mathbf H_{4,s_i^b,\bm \vartheta_{k,t}}^{(\iota)} =& \mathbf H^{c, b, (\iota)}_{\bm \varpi_{s_i^b}} \mathbf H^{\varpi} \text{, } \\
    \mathbf h_{4,s_i^b,\bm \vartheta_{k,t}}^{(\iota)} =& \mathbf z_{k,i}^{c, b} - h^{c, b}\left( \mathbf x_{ref,k}^{(\iota)}, \bm \varphi_{k,s_i^b}^{(\iota)} \right) + \mathbf H_{k, \bm \varpi_{s_i^b}}^{c, b, (\iota)} \bm \varphi_{k,s_i^b}^{(\iota)} \text{, } \\
    \mathbf H^u =& \begin{bmatrix} \mathbf I_{3 \times 3} & \mathbf 0 &\mathbf 0 \end{bmatrix} \text{, } \\
    \mathbf H^{\varpi} =& \begin{bmatrix} \mathbf 0 & \mathbf I_{3 \times 3} & \mathbf 0 \end{bmatrix} \text{, }
  \end{align}
  \end{subequations}
and $\mathbf 1_x \left( y \right)$ is an indicator function, which takes the value 1 when $x=y$, and 0 otherwise.

\end{document}